\documentclass[twocolumn,twocolappendix]{aastex701}
\usepackage{CJK}
\usepackage{amsmath}
\usepackage{enumitem}

\newcommand{\CJKJP}[1]{%
  {\begin{CJK*}{UTF8}{ipxm}#1\end{CJK*}}%
}

\newcommand{\CJKCN}[1]{%
  {\begin{CJK*}{UTF8}{gbsn}#1\end{CJK*}}%
}

\newcommand{\eg}{e.g., }
\newcommand{\ie}{i.e., }
\newcommand{\Msun}{M_{\odot}}

\begin{document}

\title{The Zero Enrichment Rare Objects (ZERO) Survey I: Photometric Calibration of NB395 and Metallicity Estimation}
\begin{CJK*}{UTF8}{gbsn}
\author[0000-0002-1259-0517]{Bowen Huang (\CJKCN{黄博闻})}
\affiliation{School of Physics and Astronomy, Beijing Normal University No.19, Xinjiekouwai St, Haidian District, Beijing, 100875, China}
\affiliation{Institute for Frontiers in Astronomy and Astrophysics, Beijing Normal University, Beijing, 102206, China}
\affiliation{National Astronomical Observatory of Japan 2-21-1, Mitaka, Tokyo 181-8588, Japan}
\affiliation{National Astronomical Observatories, Chinese Academy of Sciences, Chaoyang District, Beijing, 100101, China}
\email[show]{huangbw@nao.cas.cn}

\author[0000-0001-8537-3153]{Nozomu Tominaga (\CJKJP{冨永望})}
\affiliation{National Astronomical Observatory of Japan 2-21-1, Mitaka, Tokyo 181-8588, Japan}
\affiliation{Department of Astronomical Science, SOKENDAI (The Graduate University for Advanced Studies), 2-21-1 Osawa, Mitaka, Tokyo, 181-8588, Japan}
\affiliation{Department of Physics, Faculty of Science and Engineering, Konan University, 8-9-1 Okamoto, Kobe, Hyogo 658-8501, Japan}
\email[show]{nozomu.tominaga@nao.ac.jp}

\author[0000-0002-9053-860X]{Masashi Chiba (\CJKJP{千葉柾司})}
\affiliation{Astronomical Institute, Tohoku University, 6-3 Aoba, Sendai 980-8578, Japan}
\email[show]{chiba@astr.tohoku.ac.jp}

\author[0000-0003-2471-2363]{Haibo Yuan (\CJKCN{苑海波})}
\affiliation{School of Physics and Astronomy, Beijing Normal University No.19, Xinjiekouwai St, Haidian District, Beijing, 100875, China}
\affiliation{Institute for Frontiers in Astronomy and Astrophysics, Beijing Normal University, Beijing, 102206, China}
\email[show]{yuanhb@bnu.edu.cn}

\author[0000-0001-8424-1079]{Kai Xiao (\CJKCN{肖凯})}
\affiliation{National Astronomical Observatories, Chinese Academy of Sciences, Chaoyang District, Beijing, 100101, China}
\email{xiaokai@bao.ac.cn}

\author[0000-0002-3852-6329]{Yutaka Komiyama (\CJKJP{小宮山裕})}
\affiliation{Department of Advanced Sciences, Faculty of Science and Engineering, Hosei University, 3-7-2 Kajino-cho, Koganei, Tokyo 184-8584, Japan}
\email[]{komiyama@hosei.ac.jp}

\author[0000-0003-4656-0241]{Miho N. Ishigaki (\CJKJP{石垣美歩})}
\affiliation{National Astronomical Observatory of Japan 2-21-1, Mitaka, Tokyo 181-8588, Japan}
\affiliation{Department of Astronomical Science, SOKENDAI (The Graduate University for Advanced Studies), 2-21-1 Osawa, Mitaka, Tokyo, 181-8588, Japan}
\email[]{miho.ishigaki@nao.ac.jp}

\author[0000-0002-4013-1799]{Rosemary F. G. Wyse}
\affiliation{The William H. Miller III, Dept. of Physics \& Astronomy, Johns Hopkins University, Baltimore, MD 21218, USA}
\email[]{wyse@jhu.edu}

\author[0000-0001-6196-5162]{Evan N. Kirby}
\affiliation{University of Notre Dame, Department of Physics \& Astronomy, 225 Nieuwland Science Hall, Notre Dame, IN 46556}
\email[]{ekirby@nd.edu}

\author[0000-0001-5522-5029]{Carrie Filion}
\affiliation{Center for Computational Astrophysics, Flatiron Institute, 162 Fifth Ave, New York, NY 10010, USA}
\email[]{cfilion@flatironinstitute.org}

\author[0000-0002-8758-8139]{Kohei Hayashi (\CJKJP{林航平})}
\affiliation{National Institute of Technology, Sendai College, 4-16-1 Ayashi-Chuo, Sendai, Japan}
\email{khayashi@sendai-nct.ac.jp}

\author[0000-0002-5316-9171]{Tomoaki Ishiyama (\CJKJP{石山智明})}
\affiliation{Digital Transformation Enhancement Council, Chiba University, 1-33, Yayoi-cho, Inage-ku, Chiba 263-8522, Japan}
\email{ishiyama@chiba-u.jp}

\author[0009-0002-7874-5339]{Mikito Tanaka (\CJKJP{田中幹人})}
\affiliation{Department of Advanced Sciences, Faculty of Science and Engineering, Hosei University, 3-7-2 Kajino-cho, Koganei, Tokyo 184-8584, Japan}
\email{mikito@hosei.ac.jp} 

\author[0000-0002-3380-5302]{Hajime Susa (\CJKJP{須佐元})}
\affiliation{Department of Physics, Konan University, Okamoto, Kobe, Japan}
\email{susa@konan-u.ac.jp} 

\author[0000-0001-7679-9478]{L{\'a}szl{\'o} Dobos}
\affiliation{The William H. Miller III, Dept. of Physics \& Astronomy, Johns Hopkins University, Baltimore, MD 21218, USA}
\affiliation{E{\"o}tv{\"o}s Lor{\'a}nd University, Department of Information Systems, P{\'a}zm{\'a}ny P{\'e}ter s{\'e}t{\'a}ny 1/c, Budapest, Hungary 1117}
\email{dobos@jhu.edu}

\author[0000-0002-4108-3282]{Alexander S. Szalay}
\affiliation{The William H. Miller III, Dept. of Physics \& Astronomy, Johns Hopkins University, Baltimore, MD 21218, USA}
\affiliation{Johns Hopkins University, Department of Computer Science, 3400 N Charles Street, Baltimore, MD 21218}
\email{szalay@jhu.edu}

\author[0000-0002-7034-4621]{Tam{\'a}s Budav{\'a}ri}
\affiliation{Johns Hopkins University, Department of Applied Mathematics and Statistics, 3400 N Charles Street, Baltimore, MD 21218}
\affiliation{The William H. Miller III, Dept. of Physics \& Astronomy, Johns Hopkins University, Baltimore, MD 21218, USA}
\email{budavari@jhu.edu}

\author[0009-0008-1002-2621]{Viska Wei}
\affiliation{Johns Hopkins University, Department of Applied Mathematics and Statistics, 3400 N Charles Street, Baltimore, MD 21218}
\affiliation{Johns Hopkins University, Department of Computer Science, 3400 N Charles Street, Baltimore, MD 21218}
\email{swei20@jhu.edu}

\author[0000-0001-8239-4549]{Itsuki Ogami (\CJKJP{小上樹})}
\affiliation{National Astronomical Observatory of Japan 2-21-1, Mitaka, Tokyo 181-8588, Japan}
\affiliation{The Institute of Statistical Mathematics, 10-3 Midoricho, Tachikawa, Tokyo 190-8562, Japan}
\affiliation{Department of Advanced Sciences, Faculty of Science and Engineering, Hosei University, 3-7-2 Kajino-cho, Koganei, Tokyo 184-8584, Japan}
\email{ogami@tmu.ac.jp} 

\author[0009-0009-9769-534X]{Yoshihisa Suzuki (\CJKJP{鈴木善久})}
\affiliation{National Astronomical Observatory of Japan 2-21-1, Mitaka, Tokyo 181-8588, Japan}
\affiliation{Astronomical Institute, Tohoku University, 6-3 Aoba, Sendai 980-8578, Japan}
\email{yoshihisa.suzuki@nao.ac.jp} 

\begin{abstract}
We present the first results of the Zero Enrichment Rare Objects (ZERO) survey, a narrow-band photometric survey targeting metallicity-sensitive narrow bands using the Hyper Suprime-Cam on the Subaru Telescope. This paper outlines the scientific goals, survey strategy, current observational status, photometric calibration, and metallicity estimation. The current ZERO footprint comprises 21 pointings, covering approximately 35 deg$^2$ along the celestial equator in the NB395 band, reaching NB395 $\sim$ 24.5 mag (10$\sigma$) in one hour, significantly deeper than typical comparable narrow-band Ca H\&K surveys. The NB395 photometry is calibrated using corrected Gaia BP/RP spectra and the recalibrated SDSS Stripe 82 catalog, achieving a typical zero-point precision of $\sim$0.3 per cent and flat-field precision better than 1 per cent on spatial scales larger than 8$^\prime$. Under typical conditions (three repeats; total exposure time $\sim$1 hour), the photometric precision reaches $\sim$1.1 per cent at NB395 $\approx$ 12, $\sim$1.5 per cent at 20, and $\sim$10 per cent at 24. We estimate [Ca/H] using a Physics-constrained Stellar Locus Fitting Method. The resulting [Ca/H] precision is better than 0.18 dex for sources brighter than NB395 $\sim$ 20, spanning a wide metallicity range from metal-rich to the very metal-poor regime ($-3 \le$ [Ca/H]$\le +0.5$). We further show that NB395 traces [Ca/H] more directly than [Fe/H]. The final catalog contains one million sources with NB395 photometry, including 379,216 repeatedly observed and 61,845 with [Ca/H] estimates. These results provide a solid foundation for future ZERO survey, including photometric calibration, stellar parameter determination, and investigations of Galactic structure and chemical evolution.
\end{abstract}
\keywords{\uat{Photometry}{1234} --- \uat{Calibration}{2179} --- \uat{Metallicity}{1031} --- \uat{Chemical abundances}{224} --- \uat{Milky Way Galaxy}{1054}}

\section{Introduction} 
According to the standard cosmological model, the Universe began with the Big Bang approximately 13.8 billion years ago \citep{2020A&A...641A...6P}. Standard Big Bang nucleosynthesis produced only hydrogen, helium, and trace amounts of lithium \citep{2007ARNPS..57..463S}. Thereafter, the growth of structure in the early Universe was governed by dark matter: dark-matter halos increased their masses through hierarchical merging and accretion. Minihalos with total masses of $\sim10^6\Msun$ collapsed at $z\sim20$ -- $30$, roughly 100 million years after the Big Bang. Cooling by trace amounts of H$_2$ enabled the formation of the first stars (e.g., \citealt{1983ApJ...271..632P,1996ApJ...464..523H,1997ApJ...474....1T,1999ApJ...523L.103N,2000ApJ...544....6F,2002Sci...295...93A,2002ApJ...564...23B,2003ApJ...592..645Y}). These first stars were metal-free, \ie Population III (Pop III), stars. Because the H$_2$ cooling is relatively inefficient, their characteristic masses are expected to be more massive than $10\Msun$ (\eg \citealt{1998ApJ...508..141O,2001ApJ...561L..55O,2003ApJ...589..677O,2004ARA&A..42...79B,2006ApJ...652....6Y,2008Sci...321..669Y,2011Sci...334.1250H,Protostellar_Evolution_Hirano2014,2015MNRAS.448..568H}). At the same time, the inefficiency of the cooling favors the formation of massive, gravitationally unstable circumstellar disks, whose fragmentation can yield low-mass Pop~III stars (\eg \citealt{PopulatioIII_Clark2011a,PopulatioIII_Clark2011b,PopulatioIII_Smith2011,PopulatioIII_Greif2012,PopulatioIII_Susa2013,PopulatioIII_Susa2014}).

Massive Pop~III stars explode as supernovae a few million years after their formation, producing the first metals in the Universe (\eg \citealt{2010ApJ...724..341H,2013ARA&A..51..457N}). Second-generation stars then form from mixtures of the metals and the pristine gas. Because these stars contain metals, they are classified as Population II (Pop~II), and the fraction of low-mass stars among them can exceed that in Pop~III stars owing to the enhanced cooling efficiency provided by the metals (\eg \citealt{2001ApJ...552..464B,2002ApJ...571...30S,2021MNRAS.508.4175C}). In parallel, dark-matter halos continue to grow hierarchically. When a halo mass reaches $\sim10^8\Msun$, atomic cooling becomes efficient and the first galaxies assemble (\eg \citealt{2007ApJ...665..899W,2008ApJ...685...40W,2008MNRAS.387.1021G,2010ApJ...716..510G}). The first galaxies are therefore not metal-free and already contain second-generation
stars. The dark-matter masses of the first galaxies are comparable to those of the dwarf galaxies that surround the Milky Way in the present day \citep{2009ARA&A..47..371T,2010AdAst2010E..33R}. Continued merging of dark-matter halos drives the emergence of the earliest galaxies, now being identified by James Webb Space Telescope (JWST, \citealt{2023MNRAS.526.2542C}), and, at later times, of massive systems such as the Milky Way (\eg \citealt{2020ARA&A..58..205H}).

The evolution of the early Universe and the formation history of the Milky Way are preserved in the long-lived, low-mass stars that formed in the early Universe. Such stars have low metallicity and are referred to as metal-poor stars \citep{Beer2005AR,2015ARA&A..53..631F,2025A&ARv..33....2B}. Most Pop~III stars are unlikely to be observable in the present day, having been massive and short-lived (see \citealt{Klessen2023AR} for a recent review), whereas low-mass Pop~III or Pop~II stars can survive to the present day. In the Milky Way and nearby galaxies, individual metal-poor stars can be resolved and studied in detail. Observations provide their kinematics, chemical abundances, spatial distributions, masses, and ages. These information plays a critical role in probing the history of the early Universe and the early assembly of the Milky Way and its satellite galaxies.

The chemical abundances of metal-poor stars provide unique insights into the properties of the first stars, 
including the initial mass function \citep{1998ASPC..142...89W,2018ApJ...857...46I},
the nucleosynthetic yields of supernovae of the first stars (\eg \citealt{2007ApJ...660..516T}), 
the history of metal enrichment (\eg \citealt{2006ApJ...653.1145K}), 
and the processes that governed their own formation (\eg \citealt{2018MNRAS.475.4378C}). 
For example, the rising fraction of carbon-enhanced metal-poor (CEMP) stars toward lower metallicity suggest that carbon-based cooling played a dominant role in the early Universe (e.g., \citealt{2013ApJ...762...28N,2017MNRAS.472L.115C}). Kinematic information, dramatically enhanced by the {\tt Gaia} mission, has revealed the remnants of past galaxy mergers, \eg in the form of stellar streams, clouds, and debris \citep{2020ARA&A..58..205H,2024NewAR..9901706D,2024PASJ...76..205S}. The most significant recent merger of the Milky Way is now recognized as the Gaia-Sausage-Enceladus (GSE) \citep{2018MNRAS.478..611B,2018Natur.563...85H,2024NewAR..9901706D}. Spatial information is likewise crucial; The low-mass Pop~III
stars are predicted to concentrate on the center of halo and sub-halos \citep[e.g.][]{low-mass_Pop_III_Ishiyama2016, 2018MNRAS.474..443G}. The first
galaxies may survive as dwarf galaxies surrounding the Milky Way \citep{2009ARA&A..47..371T}. Whereas the more massive, classical dwarf galaxies have undergone substantial internal evolution, ultra-faint dwarf galaxies, many discovered in wide-area surveys such as Sloan Digital Sky Survey (SDSS, \eg \citealt{2019ARA&A..57..375S}), may have quenched early and therefore preserve the properties of ancient building blocks of the Milky Way.

The metal-poor stars are exceedingly rare: within the metal-rich disk of the Solar neighborhood, extremely metal-poor (EMP) stars with metallicity [Fe/H]\footnote{Here [A/B] $=\log_{10}(N_{\rm A}/N_{\rm B})-\log_{10}(N_{\rm A}/N_{\rm B})_\odot$, where the subscript $\odot$ refers to the solar value and $N_{\rm A}$ and $N_{\rm B}$ are the abundances of elements A and B, respectively.} ~$<-3$  occur at a frequency of approximately one in a thousand. Consequently, numerous surveys targeting metal-poor stars have been undertaken. Notable examples include Schmidt objective-prism spectroscopic surveys in the 1990s that selected candidates via the \ion{Ca}{2} H\&K lines near $395$~nm \citep{1992AJ....103.1987B,2008A&A...484..721C}. During the 2000s-2010s, wide-area,
low-resolution spectroscopic surveys employing multi-object spectrographs, such as the Sloan Extension for Galactic Understanding and Evolution (SEGUE; \citealt{SDSS_York_2000,SDSS_Yanny_2009,SDSS_Rockosi_2022}) and the Large Sky Area Multi-Object Fiber Spectroscopic Telescope (LAMOST; \citealt{LAMOST_Deng_2012,LAMOST_Liu_2014}), were carried out.

More recently, photometric surveys employing narrow-band (NB) filters sensitive to the \ion{Ca}{2} H\&K lines have gained prominence. Pioneering efforts include the SkyMapper Southern Sky Survey with $1$~m Skymapper telescope (\citealt{SMSS_Wolf_2018,SMSS_Onken_2019,SMSS_Chiti_2021,SMSSBS_Huang_2022}) and the Pristine Survey using $4$~m Canada France Hawaii Telescope (CFHT, \citealt{Pristine_Starkenburg_2017,PristineIII_Youakim2017,Pristine_Martin_2024}). NB imaging, combined with the high sensitivity of modern CCD detectors and the high cutoff accuracy  at the edge of the transmission of NB filters, offers several advantages over earlier approaches. These surveys achieve high observational efficiency by providing metallicity for all stars in a field down to the imaging depth, and the NB imaging can reach fainter magnitudes than spectroscopy. For example, the Pristine
Survey has successfully identified the C-19 stream that is proposed to be the tidal remnant of the most metal-poor globular cluster  \citep{2022Natur.601...45M}. Further programs are also performed; \eg the Javalambre Physics of the Accelerating Universe Astrophysical Survey (J-PAS; \citealt{JPAS_Benitez_2014}), the Javalambre Photometric Local Universe Survey (J-PLUS; \citealt{JPLUS_Cenarro_2019}), the Southern Photometric Local Universe Survey (S-PLUS; \citealt{SPLUS_Oliveira_2019}), the Stellar Abundances and Galactic Evolution Survey (SAGES; \citealt{SAGES_Zheng_2018,SAGES_Fan_2023}), and the Tomo-e Gozen Bright Metal-Poor Star Survey (TeMPS, \citealt{2026arXiv260827220O}).

However, the spectroscopic and photometric surveys discussed above typically reach only $\sim21$~mag \citep{Pristine_Starkenburg_2017}. This limits the accessible volume and hence studies of the outer Galactic halo and nearby dwarf galaxies. Indeed, the survey depth is crucial for the identification of low-mass Pop~III stars \citep{low-mass_Pop_III_Ishiyama2016}.
This motivates our team to develop a new NB filter sensitive to the \ion{Ca}{2} H\&K lines for the Subaru/Hyper Suprime-Cam (HSC; \citealt{HSC_Miyazaki_2018}) and conduct a deep, wide-area NB imaging survey, named the Zero Enrichment Rare Objects (ZERO) survey. HSC, mounted on an 8~m-class telescope, offers an exceptionally wide field of view and supports NB imaging. The combination of large aperture and wide field of view enables efficient, wide-and-deep mapping of the Galactic outer halo, nearby dwarf galaxies, and Galactic substructures, and offers the potential to identify low-mass Pop~III stars, thereby providing a more complete view of the Milky Way's early assembly.

This paper presents an overview of the ZERO survey, its scientific goals, observing strategy, and current status. We describe the photometric calibration of NB395, the derivation of stellar [Ca/H] using a Physics-constrained Stellar Locus Fitting Method, and the construction of the first calibrated photometric catalog. 
The paper is organized as follows.
Section 2 introduces the survey and its observations.
Section 3 presents the datasets used in this work.
Section 4 describes the photometric calibration.
Section 5 details the [Ca/H] measurement.
Section 6 summarizes the main results and outlines prospects for future studies.

\section{The Survey}

\subsection{Concept}
\begin{figure*}[htbp!]
\includegraphics[width=180mm]{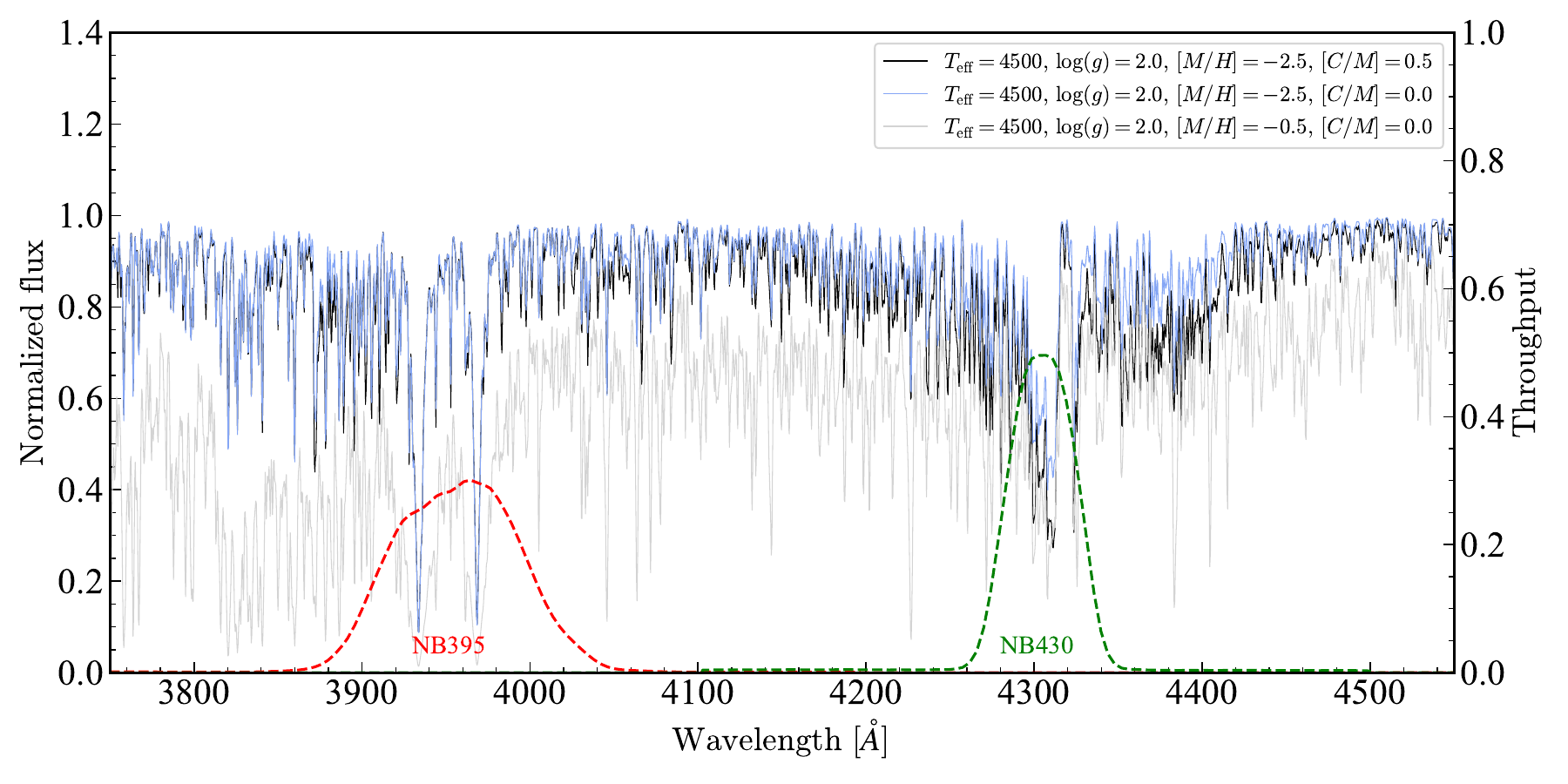}
\caption{Transmission curves of the NB395 and NB430 filters overlaid on BOSZ (\citealt{BOSZ2024}) synthetic stellar spectra with different metallicities and carbon abundances. The spectra illustrate the sensitivity of NB395 to the \ion{Ca}{2} H\&K lines and of NB430 to the CH molecular ($G$-band) features.}
\label{filter}
\end{figure*}

The ZERO survey is a deep, wide-area narrow-band program targeting metal-poor and potentially metal-free stars in the Galactic halo and nearby dwarf galaxies (PI: M. Chiba). It employs Subaru/HSC, a highly efficient wide-field imager, together with two custom narrow-band filters: NB395, sensitive to the \ion{Ca}{2} H\&K lines, and NB430, sensitive to the CH molecular ($G$-band) features. Figure~\ref{filter} shows the transmission curves of the two filters overlaid on synthetic stellar spectra with different metallicities and carbon abundances, illustrating their sensitivity to these spectral features. The filter development team for the ZERO survey defined the central wavelengths and widths of filter transmissions to maximize the ability to photometrically measure the metallicity\footnote{We following adopt [Ca/H] as a indicator of metallicity instead of [Fe/H] because the NB395 filter is sensitive to \ion{Ca}{2} lines (see section 5).} and the carbon abundance. The NB395 and NB430 filters became available in Aug 2021 and Feb 2020, respectively. The transmission curves at the center and the area-weighted mean were measured in the laboratory by \cite{2018PASJ...70...66K}.\footnote{\url{https://subarutelescope.org/Instruments/HSC/sensitivity.html}}

With one hour of integration, the $10\sigma$ depths in the NB395 and NB430 bands reach $24.5$ and $24.4$ mag, respectively. For typical colors of metal-poor stars, this corresponds to $G\sim22.5-23.4$~mag. At these depths, red-giant stars are detectable out to $\sim1$~Mpc, while dwarf stars can be detected to $\sim100$~kpc. Consequently, the survey reaches the Galactic outer halo and the dwarf galaxies surrounding the Milky Way and tests the existence of low-mass Pop~III stars.

The survey fields are located within the wide layer of the HSC Subaru Strategic Program (HSC-SSP; \citealt{HSCSSP_Arihara_2018a,HSCSSP_Arihara_2018b,HSCSSP_Arihara_2019,HSCSSP_Arihara_2022}), for which broadband $g$- and $i$-band imaging data are already available. We employ two narrow-band filters, NB395 and NB430, in combination with broadband photometry from the HSC-SSP.
Since spectroscopy of large numbers of faint sources is challenging, photometric metallicities provide a powerful alternative to characterize metallicities of faint populations over a wide area, as demonstrated in previous studies (e.g. \citealt{2025ApJ...989...48H,2026arXiv260203014O}). Considering the high fraction of CEMP stars at low metallicity, the determination of photometric carbon abundances is also important to constrain the formation mechanisms of metal-poor stars and study the evolution of the early Universe. Further, the combination between photometric metallicities and photometric carbon abundances are crucial to identify genuinely metal-free stars that exhibit neither \ion{Ca}{2} nor CH absorption. 

The aims of the ZERO survey are to
\begin{enumerate}
 \item Discover pristine, low-mass Pop~III stars with $\le 0.8 M_\odot$ or place stringent constraints on their existence;
 \item Identify metal-poor stars in the Galactic outer halo and in dwarf galaxies;
 \item Constrain the low-metallicity tail of the metallicity distribution function (MDF) in both the Milky Way and its dwarf galaxies;
 \item Constrain the initial mass function (IMF) of the first stars;
 \item Constrain the formation sites of metal-poor stars;
 \item Investigate the evolution of the IMF with metallicity;
 \item Provide metal-poor targets for ongoing spectroscopic
       observations with, \eg Subaru/Prime Focus Spectrograph \citep{2014PASJ...66R...1T,2016SPIE.9908E..1MT,2026arXiv260409875C} and future 30m-class telescopes.
\end{enumerate}

\subsection{Observations and coverage}
The pilot observations of the ZERO survey were conducted on multiple nights between September and October in 2021 (PI: M. Chiba, ID: S21B-102). Observations were carried out using the Subaru HSC, with a combination of second-half night observations on September 3rd and 4th, a full-night observation on October 1st, and second-half night observations on October 2nd. 
Dome flats for the NB395 observations were obtained once during each observing run (\citealt{HSC_Pipe_Bosch2018}), where an observing run corresponds to an approximately two-week dark-moon observing period. Thus, the same dome flats were used for the September 3--4 observations, and another set was used for the October 1--2 observations.
The field layout defined for the pilot observations (R.A. = $-30^\circ$ $\sim$ $40^\circ$, Decl. $\sim$ $0^\circ$) spans the HSC-SSP Wide Fall equatorial field region, with individual footprints placed at intervals of 1.5 degrees. These footprints are labeled as \texttt{ZERO\_1} through \texttt{ZERO\_47},whereas \texttt{ZERO\_42C} and \texttt{FEIGE110} lie slightly off the celestial equator. The observed footprint in the pilot observations is presented in Table \ref{FP} excluding standard-star fields such as \texttt{FEIGE110}.

\begin{table}[htbp!]
\centering
\setlength{\tabcolsep}{2pt}
\caption{The observed footprint of the ZERO survey.\label{FP}}
\begin{tabular}{lcccc}
\hline
Field & R.A. (deg) & Decl. (deg) & l (deg) & b (deg) \\
\hline
\texttt{ZERO\_1} & $-$30.000 & $+$0.000 & $+$59.101 & $-$40.721 \\
\texttt{ZERO\_2} & $-$28.500 & $+$0.000 & $+$60.314 & $-$41.913 \\
\texttt{ZERO\_13} & $-$12.000 & $+$0.000 & $+$77.474 & $-$53.855 \\
\texttt{ZERO\_14} & $-$10.500 & $+$0.000 & $+$79.486 & $-$54.790 \\
\texttt{ZERO\_15} & $-$9.000 & $+$0.000 & $+$81.590 & $-$55.690 \\
\texttt{ZERO\_16} & $-$7.500 & $+$0.000 & $+$83.792 & $-$56.552 \\
\texttt{ZERO\_25} & $+$6.000 & $+$0.000 & $+$108.153 & $-$62.082 \\
\texttt{ZERO\_26} & $+$7.500 & $+$0.000 & $+$111.306 & $-$62.387 \\
\texttt{ZERO\_27} & $+$9.000 & $+$0.000 & $+$114.516 & $-$62.619 \\
\texttt{ZERO\_28} & $+$10.500 & $+$0.000 & $+$117.769 & $-$62.777 \\
\texttt{ZERO\_35} & $+$21.000 & $+$0.000 & $+$140.349 & $-$61.766 \\
\texttt{ZERO\_36} & $+$22.500 & $+$0.000 & $+$143.363 & $-$61.333 \\
\texttt{ZERO\_37} & $+$24.000 & $+$0.000 & $+$146.290 & $-$60.835 \\
\texttt{ZERO\_38} & $+$25.500 & $+$0.000 & $+$149.122 & $-$60.275 \\
\texttt{ZERO\_41} & $+$30.000 & $+$0.000 & $+$157.005 & $-$58.262 \\
\texttt{ZERO\_42c} & $+$31.325 & $-$4.275 & $+$163.806 & $-$61.141 \\
\texttt{ZERO\_43} & $+$33.000 & $+$0.000 & $+$161.742 & $-$56.675 \\
\texttt{ZERO\_44} & $+$34.500 & $+$0.000 & $+$163.958 & $-$55.818 \\
\texttt{ZERO\_45} & $+$36.000 & $+$0.000 & $+$166.077 & $-$54.923 \\
\texttt{ZERO\_46} & $+$37.500 & $+$0.000 & $+$168.102 & $-$53.993 \\
\texttt{ZERO\_47} & $+$39.000 & $+$0.000 & $+$170.036 & $-$53.031 \\
\hline
\end{tabular}
\end{table}

\begin{figure*}[htbp!]
\includegraphics[width=180mm]{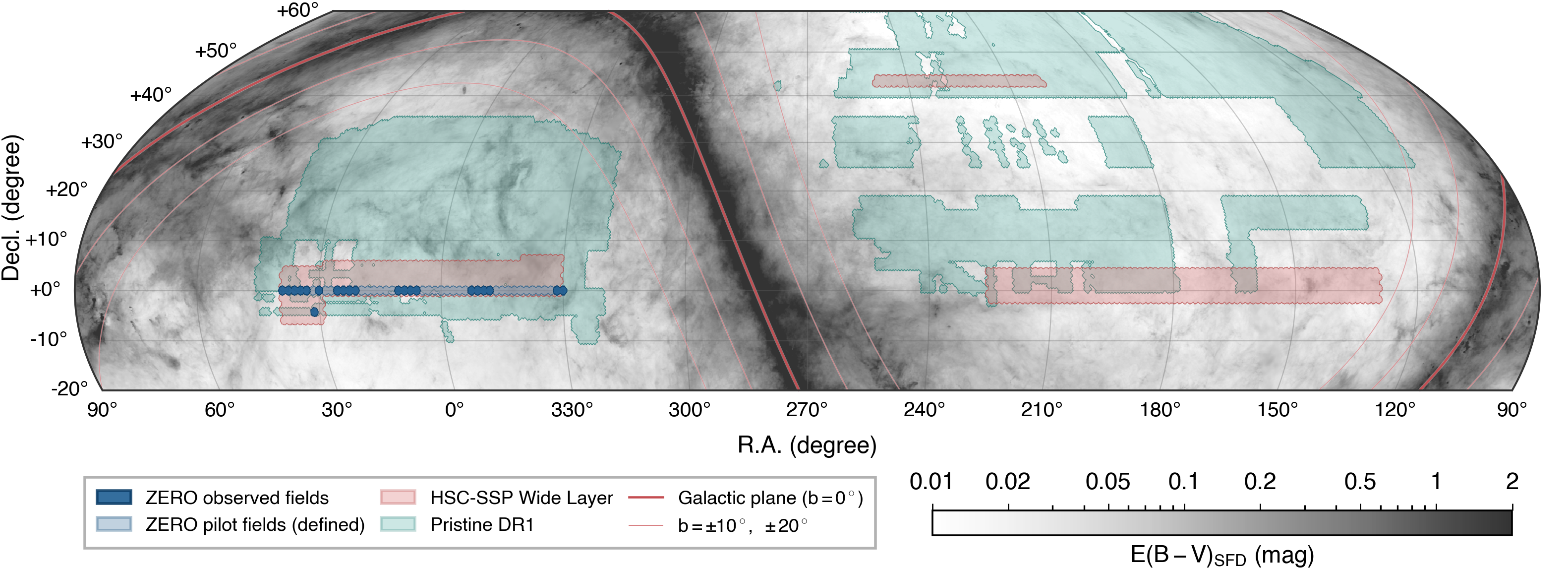}
\caption{Sky coverage of the ZERO survey, together with the HSC-SSP Wide Layer and Pristine DR1 footprints. The field layout defined for the pilot observations and the observed fields are shown separately. The background shows the SFD98 dust map, with the Galactic plane indicated for reference.}
\label{FootPrint}
\end{figure*}

Figure~\ref{FootPrint} shows the sky coverage of the ZERO survey together with the HSC-SSP Wide Layer (\citealt{HSCSSP_Arihara_2018a,HSCSSP_Arihara_2018b,HSCSSP_Arihara_2019,HSCSSP_Arihara_2022}) and Pristine DR1 footprints. The field layout defined for the pilot observations extends approximately along the celestial equator from R.A. = $-30^\circ$ to $40^\circ$, overlapping with the HSC-SSP Wide Layer, which provides the broadband photometry used in this work. The observed fields are highlighted separately. These fields were defined specifically for the pilot observations and do not imply that subsequent ZERO observations will follow the same field layout. The Pristine DR1 (\citealt{Pristine_Martin_2024}) footprint is also shown for comparison, together with the Galactic plane and the two-dimensional dust map of \cite{SFD_1998} (hereafter, SFD98) in the background.

For each observation, footprints with an elevation range of $65 \sim 70^\circ$ were selected to optimize observing conditions. Each pointing was observed for a total of 1 hour, consisting of one short 30-second exposure for calibration purposes followed by three long exposures of 1170 seconds each. The survey covers the region also overlaps with the Pristine survey field, making it a crucial area for comparison between the HSC/NB395 and Pristine Ca H\&K data.
While the data collected so far consists of 21 fields covering approximately 35 deg$^2$ and these fields are not contiguous, we plan to cover a wider area in the full survey.

\subsection{Preliminary data processing}

We perform CCD processing with the standard reduction pipeline of HSC ({\tt hscPipe} v8.5.3, \citealt{HSC_Pipe_Bosch2018}). The {\tt hscPipe} provides full packages for data analyses of images obtained with HSC\@. Since we apply the photometric calibration dedicated for the narrow-band photometry (Section~\ref{sec:photocal}), we apply only the bias subtraction, flat-fielding, global background subtraction, astrometry, source detection, and source measurement for the CCD processing. The astrometric calibration is made relative to the PS1 catalog with a $4.0$ arcsec ($24$ pixel) aperture diameter. The astrometric calibration fails for those CCD images in which only a small number of sources are detected. This frequently occurs for the images with short exposure time. The failed CCD images are excluded from the following analysis.

\section{Datasets}
In this section, we provide a detailed description of datasets used in the photometric calibration and the metallicity estimation.
\subsection{Gaia Data Release 3}
The ESA Gaia mission aims to obtain precise astrometric, photometric, and spectroscopic measurements for stars in the Milky Way and the Local Group (\citealt{Gaia_col_2016}). Gaia Data Release 3 (DR3) provides astrometry for about 1.47 billion sources with full five-parameter solutions (positions, parallax, and proper motions), and broadband $G$, $BP$, and $RP$ photometry for more than 1.8 billion objects (\citealt{GaiaDR3Summary_2023a}). 

In addition, DR3 includes approximately 220 million low-resolution BP/RP (XP) spectra covering 336 -- 1020 nm, from which synthetic photometry (XPSP) can be generated for arbitrary bands within this wavelength range (\citealt{GaiaDR3XPSP_2023b}).
Instead of providing flux as a function of wavelength, these data are expressed as projections on 55 orthonormal Hermite basis functions for both BP and RP channels (\citealt{GaiaXPInternalCalibration_Carrasco_2021,GaiaXPProcessingValidation_Angeli_2023,GaiaXPExternalCalibration_Montegriffo_2023}).
However, systematic trends have been identified in the XP spectra, depending on color, magnitude, and interstellar extinction (\citealt{GaiaXPExternalCalibration_Montegriffo_2023,XPcor_Huang_2024a,XPCor_Ye_2025}). \cite{XPcor_Huang_2024a} developed wavelength-domain corrections for these effects, valid for $-0.5 < BP-RP < 2$, $3 < G < 17.5$, and $E(B-V) < 0.8$. Validation showed that these corrections effectively reduce systematic errors, reaching an internal precision of about 1 -- 2 per cent.

\subsection{Recalibrated SDSS Stripe 82 Standard Stars}
The repeatedly scanned Stripe 82 region ($|\mathrm{Decl.}| < 1.266^\circ$, $-51.5^\circ < \mathrm{R.A.} < 60^\circ$) is a contiguous equatorial area of about 300 deg$^2$ in the SDSS\@. \cite{S82_Ivezic_2007} delivered an accurate photometric catalog of about one million stars in the $u$, $g$, $r$, $i$, and $z$ bands, achieving an internal calibration consistency of approximately one per cent. The random photometric errors are below 0.01 mag for stars brighter than 19.5, 20.5, 20.5, 20.0, and 18.5 mag in the $u$, $g$, $r$, $i$, and $z$ bands, respectively, thereby providing a practical definition of the SDSS photometric system.

In this study, we adopt the recalibrated version of the v4.2 catalog of Stripe 82 (\citealt{S82v42}) presented by \cite{RS82_Huang_2022a} (hereafter RS82). The original v4.2 catalog provides 2-3 times more measurements per star, with random errors 1.4-1.7 times smaller than those in \cite{S82_Ivezic_2007}.
The RS82 catalog identifies and corrects large-scale, correlated zero-point errors present in the original SDSS Stripe 82 photometry. The recalibration reveals systematic offsets of approximately 2.5 per cent in the $u$ band and about 1 per cent in the $griz$ bands. After correction, the internal precision of RS82 improves to about 5 mmag in $u$ and 2 mmag in $griz$, representing an improvement of roughly a factor of five compared to the original catalog.

\subsection{SDSS SEGUE}
The Sloan Digital Sky Survey Data Release 12 (SDSS DR12; \citealt{SEGUE_Alam_2015}), the final release of the SDSS-III project, includes all spectroscopic observations obtained up to July 2014. Stellar spectra from SDSS are analyzed using the SEGUE Stellar Parameter Pipeline (\citealt{SEGUEPipe_Lee_2008a,SEGUEPipe_Lee_2008b,SEGUEPipe_Smolinski_2011}), which provides three fundamental stellar parameters—effective temperature ($T_{\mathrm{eff}}$), surface gravity (log $g$), and metallicity ([Fe/H]) -- for most stars within the temperature range of 4000 -- 10,000 K and with spectral signal-to-noise ratios greater than 10. The SEGUE is adopted as an auxiliary spectroscopic reference in the calibration process.

\subsection{HSC-SSP Wide layer}
The Hyper Suprime-Cam Subaru Strategic Program (HSC-SSP) is a wide and deep field imaging survey carried out using the HSC\@. The survey was conducted from March 2014 to January 2022, spanning approximately 7.5 years (\citealt{HSCSSP_Arihara_2018a,HSCSSP_Arihara_2018b,HSCSSP_Arihara_2019,HSCSSP_Arihara_2022}), and collected photometric data in multiple optical bands with varying depths over a large fraction of the sky. In this work, we utilize only the Wide layer for stellar metallicity estimation. The Wide layer currently covers about 670 deg$^2$ and will encompass about 1100 deg$^2$ in the final data release in the g, r, i, z, and y bands, with 5-sigma limiting magnitudes of approximately 26.5, 26.5, 26.2, 25.2, and 24.4, respectively (\citealt{HSCSSP_Arihara_2022}). 
This catalog was generated by the {\tt hscPipe} v8.4, updated with an FGCM code ({\tt hscPipe} v8.5.3), based on the pipeline developed for the Vera~C.~Rubin Observatory (\citealt{2008SerAJ.176....1I,2017ASPC..512..279J,2019ASPC..523..521B}) and calibrated against Pan-STARRS1 photometry and astrometry (\citealt{2012ApJ...756..158S,2012ApJ...750...99T,2013ApJS..205...20M,2016arXiv161205560C}).

\subsection{BEST Standard stars in the HSC-SSP Photometric System}
The BEst STars (BEST) database (\citealt{BEST_Xiao_2026}) contains more than 200 million XPSP standard stars, constructed using an updated version of the XPSP methodology (\citealt{XPSP_Xiao_2023c}) applied to the corrected Gaia XP spectra (\citealt{XPcor_Huang_2024a}). We adopt the transmission curves of the HSC-SSP photometric system\footnote{\url{https://subarutelescope.org/Instruments/HSC/sensitivity.html}} to generate the synthetic magnitudes in NB395, g, r and i bands. A detailed description of the XPSP method and the construction of the standard stars is provided in Section 3.2 of \cite{XPSP_Xiao_2023c}.
These standard stars are distributed across the entire sky and are reported in the AB magnitude system (\citealt{AB_Oke_1983}).

\subsection{DD-Payne for LAMOST DR9}
We make use of stellar atmospheric parameters and elemental abundances from the LAMOST DR9 catalog derived with the data-driven Payne method (DD-PAYNE; \citealt{LMDDP_Zhang_2025a}) as a spectroscopic reference for metallicity estimation. The catalog provides homogeneous measurements of effective temperature, surface gravity, and 22 individual elemental abundances for millions of stars based on low-resolution LAMOST spectra. In this catalog, the [Fe/H] measurements are reliable down to $\sim-4.0$, while most elemental abundance ratios [X/Fe] are robust for stars with [Fe/H] $\ge-2.0$. For high signal-to-noise spectra (S/N $>$ 50), typical uncertainties reach $\sim$ 0.05 dex for light elements, with measurement errors that depend strongly on spectral quality and scale approximately inversely with S/N.

\section{Photometric Calibration of NB395} \label{sec:photocal}
\subsection{Overview}
The current dataset includes observations of 21 fields, comprising a total of 89 visits (exposures). Among them, 58 visits have an exposure time of 1170 s, and 21 visits have an exposure time of 30 s. In addition, 7 visits have an exposure time of 600 s, mainly covering \texttt{ZERO\_44} and \texttt{ZERO\_47} fields. Several other visits were terminated prematurely due to weather or technical issues, resulting in unreliable photometric quality.

We first remove measurements affected by saturation, blending, or bad pixels, as well as those located near the CCD edges, and measurements with signal-to-noise ratios below 1. For sources with  $G<21$, whose $NB395$ magnitudes are typically brighter than 22 -- 24, the vast majority have counterparts with Gaia DR3 photometry, identified via the Gaia coordinates and \texttt{source\_id}. This subset contains 113,314 sources and a total of 330,695 individual observations. 
For photometry without Gaia counterparts within 1.5 arcsec, we identify repeated observations based on positional proximity in equatorial coordinates (within 5 arcsec) and similarity in the calibrated $NB395$. This yields 906,280 unique sources and 1,313,111 individual observations, among which 288,834 sources have at least two repeated observations, while the remainder are detected only once. 

Given these characteristics, we construct the following photometric calibration model, which performs an independent calibration for each visit and is applied to each observation. The model is expressed as:
\begin{equation}
\begin{aligned}
    NB395_{\mathrm{cali}} &= NB395_{\mathrm{uncali}} + ZP(i) + SS(i,n,nX,nY) \\
    &\quad + ZP_{\mathrm{Blue/Red}}(i)
\end{aligned}
    \label{EQ1}
\end{equation}
where $NB395_{\mathrm{cali}}$ and $NB395_{\mathrm{uncali}}$ represent the calibrated and uncalibrated NB395 magnitudes of stars in each visit, respectively. The term $ZP(i)$ denotes the zero-point offset for each visit $i$, $SS(i,n,nX,nY)$ describes the spatial structure within the FoV, where $n$ indicates the CCD number on HSC, while $nX$ and $nY$ are the relative positions with respect to the CCD center, and the $ZP_{Blue/Red}(i)$ indicates the additional zero-point offsets applied for blue and red stars, defined by $BP-RP < 1.25$ and $BP-RP > 1.25$, respectively, in each visit. However, the color information required for $ZP_{Blue/Red}(i)$ is from Gaia photometry. Consequently, photometric measurements in the ZERO survey that lack Gaia counterparts do not include this correction term. The amplitude of this correction is typically smaller than 1 per cent. Since sources without Gaia counterparts are generally faint, their photometric uncertainties are substantially larger than this level, and thus the omission of this correction is not expected to introduce significant systematic effects.
For each $SS(i,n,nX,nY)$, the spatial structure is modeled as a second-order two-dimensional polynomial with respect to the stellar position $(nX,nY)$:
\begin{equation}
\begin{aligned}
SS(i,n,nX,nY) &= a_1(i,n)nX^2 + a_2(i,n)nX \\
              &\quad + a_3(i,n)nY^2 + a_4(i,n)nY \\
              &\quad + a_5(i,n)nX\cdot nY + a_6(i,n) \\
\end{aligned}
\label{EQ2}
\end{equation}

All parameters in the above calibration model are determined by standard stars. Although the constant term $a_6(i,n)$ is formally degenerate with the visit $ZP(i)$ in Equation \ref{EQ1}, these terms are solved sequentially during the iterative procedure, where $ZP(i)$ first characterizes the global offset of each visit and $SS(i,n,nX,nY)$ subsequently models the relative spatial variations in the FoV.

The core of the photometric calibration is to build a sufficiently large set of NB395 narrow-band standard stars to constrain the model parameters. The most direct method is to derive NB395 photometry from Gaia XP spectra via synthetic photometry (XPSP) using the HSC NB395 filter curve. However, because Gaia XP spectra are limited to $G < 17.65$, only a few bright sources are covered in the 1170s exposures under the current ZERO survey strategy. These sources are also unevenly distributed across the FoV, and some fields show stripe-like gaps in the Gaia XP coverage, likely due to Gaia's scanning law.
The RS82 catalog, in contrast, supplies a dense set of faint stars that fully covers the equatorial region of the current ZERO survey. As shown by the Stellar Color Regression method (SCR; \citealt{SCR_Yuan_2015,RS82_Huang_2022a}), RS82 $ugi$ photometry can be transformed into the NB395 system, yielding a large, well-distributed standard-star set to constrain the calibration parameters.
For fields away from the celestial equator, the zero point $ZP(i)$ is derived from the limited XPSP standard stars, while the spatial structure term $SS(i,n,nX,nY)$ is taken directly from temporally adjacent equatorial visits.

\begin{figure}[htbp!]
\centering
\includegraphics[width=\columnwidth,height=0.8\textheight,keepaspectratio]{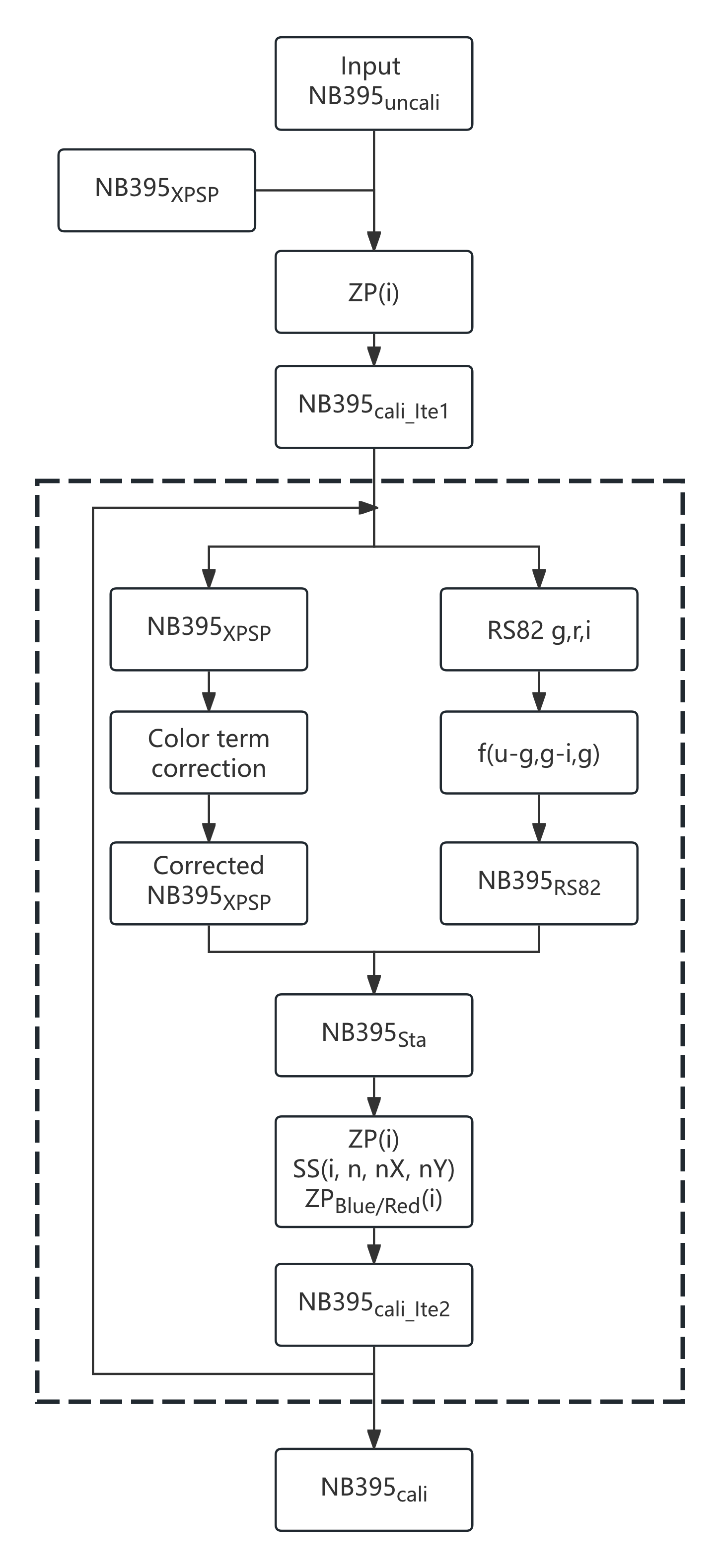}
\caption{Flowchart of the calibration procedure. The steps enclosed by the dashed box are iterated once more.} 
\label{Flow}
\end{figure}

The full calibration procedure is shown in Figure~\ref{Flow}. We first constrain the zero-point using XPSP standard stars, then construct a transformation relation from RS82 sources. All calibration parameters are then jointly constrained with both datasets, and the procedure is iterated to ensure a robust transformation and calibration.

\subsection{Data Selection}
There are approximately 17,000 common sources (about 70,000 measurements) matched with Gaia XP spectra, and about 70,000 common sources (180,000 measurements) matched with RS82.
To ensure the reliability of the HSC photometry, the calibration models, and the transformation relation from the SDSS system to NB395, the following selection criteria are applied to the ZERO, XPSP, and RS82 datasets:

\begin{enumerate}[itemsep=0.2cm, topsep=0.5cm, parsep=0.2cm, partopsep=0.3cm]
    \item For the HSC photometry, aperture photometry with an fixed-aperture radius of 12 pixels (2 arcsec) are adopted. 
    \item For the Gaia XP spectra, sources with large $\texttt{phot\_bp\_rp\_excess\_factor}$ values or those significantly deviating from the general relation between $\texttt{phot\_bp\_rp\_excess\_factor}$ and the $BP-RP$ color are removed. 
    Specifically, sources satisfying
    \begin{equation*}
    \begin{aligned}
    \texttt{phot\_bp\_rp\_excess\_factor} > \\
     -0.020(BP-RP)^3 + 0.135(BP-RP)^2 \\
     - 0.133(BP-RP) + 1.232
    \end{aligned}
    \end{equation*}
    are excluded.
    The $\texttt{phot\_bp\_rp\_excess\_factor}$ is defined as $C = \frac{I_{BP} + I_{RP}}{I_G}$, which evaluates background and contamination effects in both $BP$ and $RP$ photometry and spectra. This empirical criterion helps exclude sources with low-quality Gaia measurements and effectively removes potential extended sources such as galaxies.
    \item For RS82, we restrict the color range to $0.25 < g-i < 3$. To ensure sufficient faint standard stars, only a small number of sources with standard errors that are significantly greater than the main magnitude-error relation are excluded. Additionally, we require \texttt{u\_msig} $< 0.5$.
\end{enumerate}

\subsection{Standard Stars}
As mentioned earlier, our standard stars consist of two parts, namely XPSP and RS82. This subsection focuses on the generation of the standard stars and the practical definition of the $NB395$ band.

In wide-field photometric observations, the transmission curve of a filter can vary with the incident angle and the position within the FoV. Without auxiliary measurements, these variations are usually difficult to be determined precisely. Therefore, when photometric calibration relies entirely on standard stars, the photometric band must be clearly defined. 
Although the mean transmission curve of the $NB395$ band is measured in the laboratory, small discrepancies on the mean transmission curve of the $NB395$ band may potentially exist between the laboratory characterization and the in-situ performance. Consequently, in the pilot survey, the definition of $NB395$ is ultimately determined precisely and implicitly by the photometric observations rather than explicitly by the laboratory measurements, although the two are in close agreement.

For the XPSP sample, synthetic photometry is performed using Gaia XP spectra combined with the area-weighted mean transmission curve of the HSC $NB395$ filter. Due to systematic errors in Gaia XP spectra and possible discrepancies between the adopted and the actual transmission curves, a color term may exist between $NB395_{\mathrm{XPSP}}$ and the observed $NB395$. This color term is assumed to be linear and corrected during each calibration iteration, ensuring that the mean $NB395_{\mathrm{XPSP}}$ values remain consistent with the observed $NB395$.

For the RS82 sample, we apply the least-squares boosting (LSBoost) regression method to establish the empirical relation $NB395 = f(u-g, g-i, g)$ based on the Step 1 result $NB395_{\mathrm{cali\_Ite1}}$, and then generate $NB395_{\mathrm{RS82}}$ using the RS82 $ugi$ magnitudes.
The final empirical relation after the iterative fitting procedure is illustrated at a representative magnitude of $g=19$ in Figure~\ref{RS82_relation} in Appendix B.
In this relation, $f$ implicitly incorporates the mean transmission curve of the $NB395$ band derived from actual observations, meaning that $NB395_{\mathrm{RS82}}$ is defined with respect to the observed mean transmission curve.

\begin{figure*}[htbp!]
\includegraphics[width=180mm]{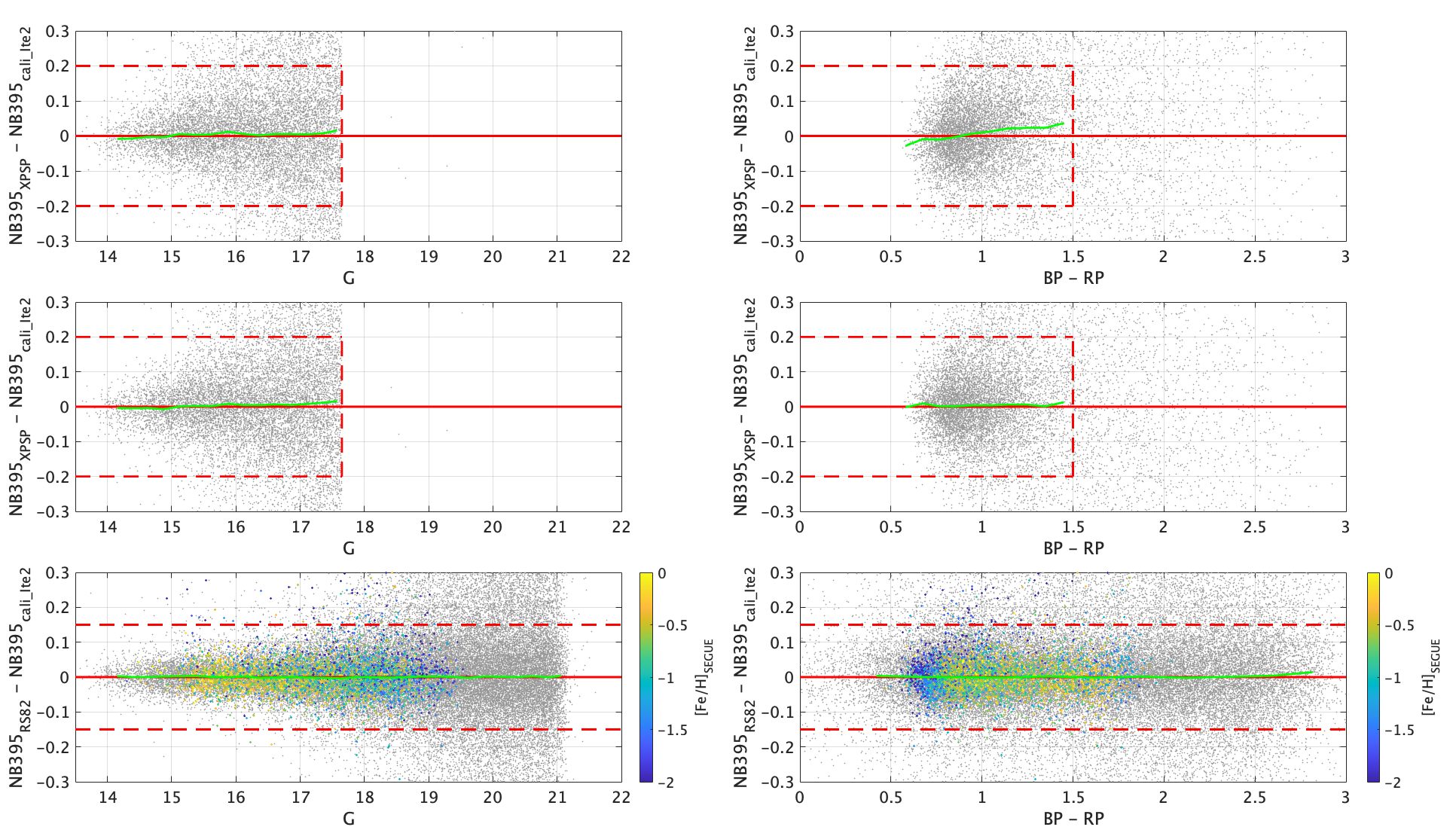}
\caption{Differences of the XPSP (top), color-term-corrected XPSP (middle) and RS82 (bottom) standard stars relative to $NB395_{\mathrm{cali\_Ite2}}$, shown as functions of Gaia $G$ magnitude and $BP-RP$ color. The color of dots in the bottom panels indicates [Fe/H] values from SDSS SEGUE. The green curve in each panel indicates the median trend. The solid horizontal line denotes zero difference, while the dashed lines indicate the adopted selection limits for the standard stars. Only sources within the dashed lines are included in the standard-star sample.} 
\label{Standard_Star}
\end{figure*}

The comparison between the standard stars $NB395_{\mathrm{XPSP}}$ and $NB395_{\mathrm{RS82}}$ and the observed $NB395_{\mathrm{cali\_Ite2}}$ from the final iteration is shown in Figure \ref{Standard_Star}. The XPSP sample is limited to $G < 17.65$, with uncertainties increasing toward fainter and redder stars. In contrast, the RS82 results agree well with the observations, with no significant systematic differences between metal-rich and metal-poor stars. This confirms the reliability of the $NB395 = f(u-g, g-i, g)$ transformation. To ensure robust model parameter constraints in the following analysis, we remove sources with $BP-RP > 1.5$ from the XPSP sample. We then require $|NB395_{\mathrm{XPSP}} - NB395_{\mathrm{cali\_Ite2}}| < 0.20$ for XPSP and adopt the stricter condition $|NB395_{\mathrm{RS82}} - NB395_{\mathrm{cali\_Ite2}}| < 0.15$ for RS82.

In the subsequent photometric calibration procedure, the XPSP and RS82 standard star samples are combined. Considering the higher precision and larger sample size of RS82, for sources common to both catalogs, only the RS82 results are retained, with XPSP serving merely as a supplementary dataset. Therefore, within the equatorial region, RS82 standard stars dominate the constraints on the calibration model parameters. 
The sample of NB395 magnitudes derived from the XPSP and RS82 standard stars is hereafter referred to as $NB395_{\mathrm{Sta}}$.

\subsection{Zero-Points and Spatial Structures of Visits}
\begin{figure*}[htbp!]
\includegraphics[width=180mm]{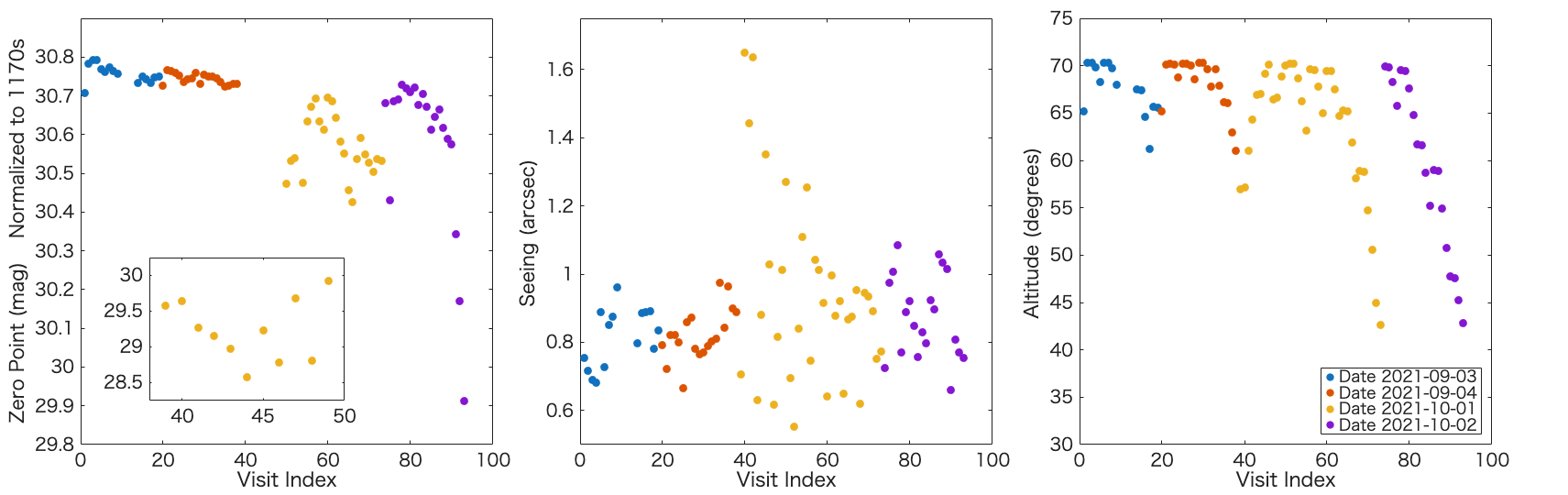}
\caption{Overview of zero-points, seeing, and altitude for all visits over the four nights.}
\label{ZP}
\end{figure*}

The final-iteration zero-points $ZP(i)$ (Figure\,\ref{ZP}) are mainly determined by atmospheric conditions and visit altitude (air mass). On the first half of the night of October 1st, persistent thin clouds caused noticeable variations in both seeing and zero-point. On stable nights such as September 3rd and 4th, zero-point variations at high altitude are typically below $0.05$-$0.06$ mag, with seeing remaining stable. In contrast, on poorer nights such as October 1st and 2nd, zero-point variations are larger.

\begin{figure*}[htbp!]
\includegraphics[width=180mm]{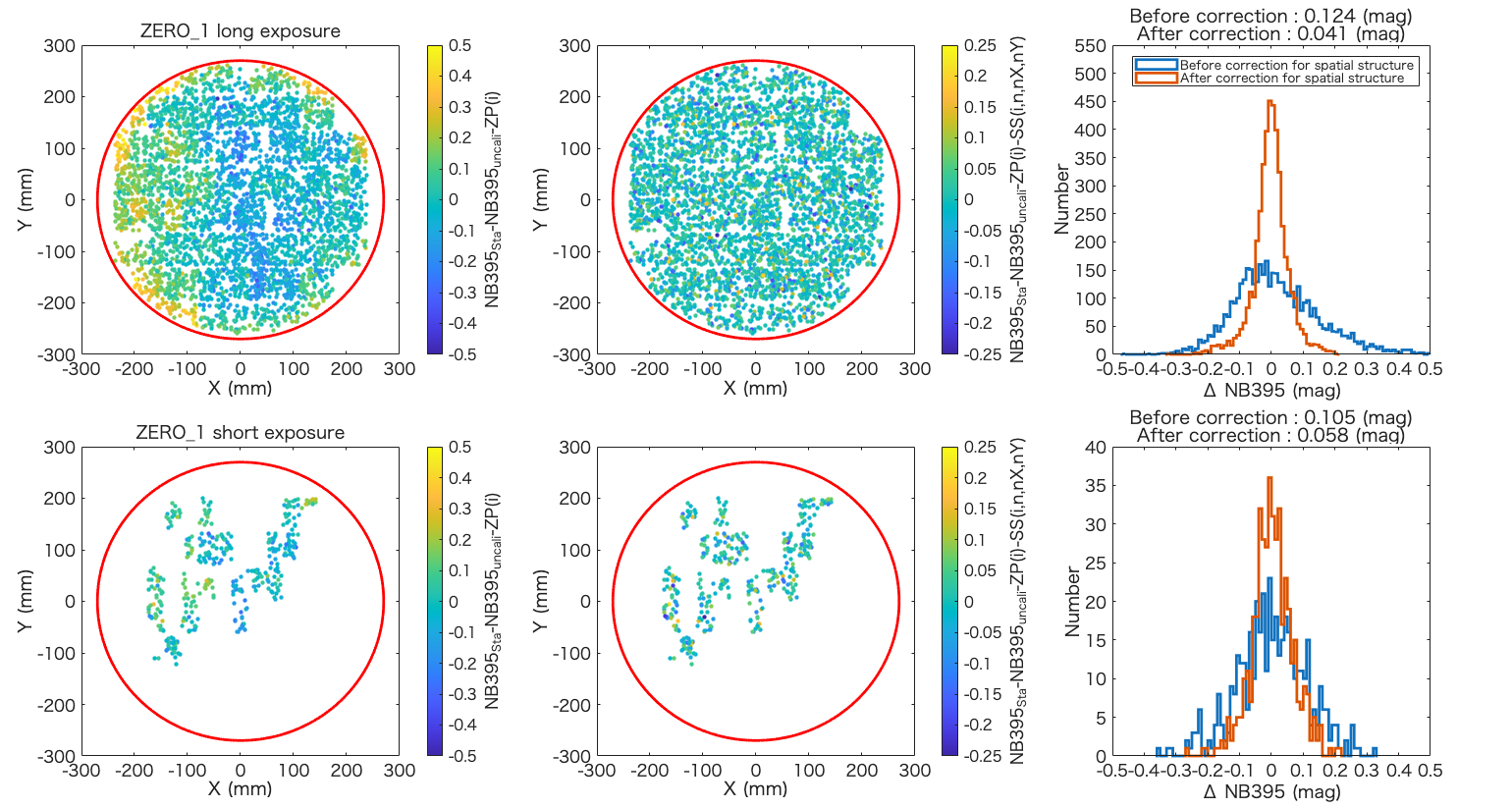}
\caption{Typical spatial structures and their corrections for a long-exposure visit and a short-exposure visit of field \texttt{ZERO\_1}. The first column shows the spatial distribution of the differences between the observed $NB395$ and the standard star $NB395_{\mathrm{Sta}}$ before spatial structure correction. The second column presents the residuals after correction, while the third column compares the calibrated magnitudes with the standard stars before and after this correction.}
\label{SS}
\end{figure*}

For the spatial structure component $SS(i,n,nX,nY)$, Figure \,\ref{SS} shows spatial structures on the FoV for typical long- and short-exposure visits of field \texttt{ZERO\_1}. Additional examples from the four observing nights are presented in Appendix C. In the long exposure, a clear ring-like pattern with amplitudes up to $\pm0.4$ mag and discontinuous zero-point jumps between CCDs are visible across the FoV. Despite the limited number of CCDs with valid readouts in the short exposure, a similar pattern appears. Figure \,\ref{SS} also shows that our correction effectively removes these structures. After correction, the dispersion between $NB395_{\mathrm{cali}}$ and the standard $NB395_{\mathrm{Sta}}$ is about 0.041 mag for long exposures and 0.058 mag for short exposures (likely due to the smaller number of standard stars available in each CCD, which leads to larger random uncertainties in the corrections).

\subsection{Color dependent Zero-Points and Spatial Structures}
\begin{figure*}[htbp!]
\includegraphics[width=180mm]{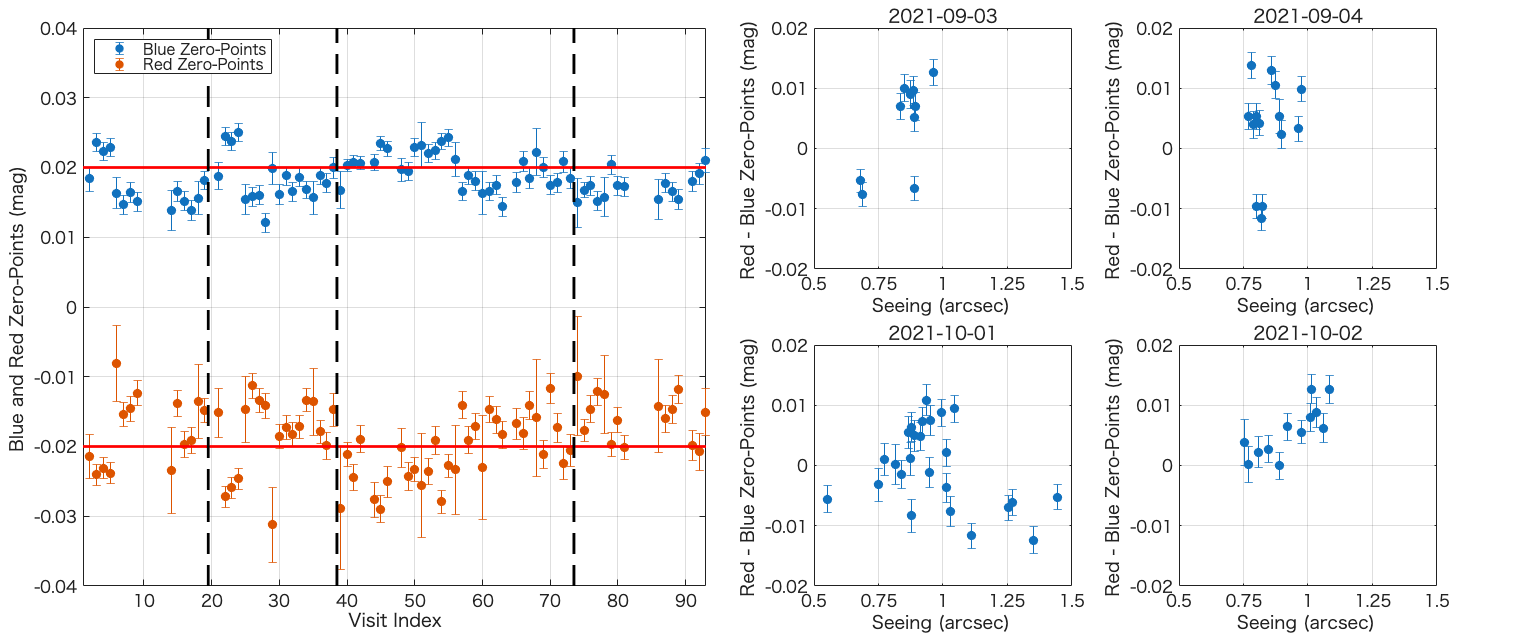}
\caption{The left panel shows the blue- and red-star zero-points for each visit, shifted upward and downward by 0.02 mag, respectively. The error bars represent the uncertainties of the zero-points, estimated from the number of available standard stars. The black vertical dashed lines separate the four observing nights. The four panels on the right illustrate, for each night, the relation between the seeing and the zero-point differences between the blue and red stars.}
\label{Color_ZP}
\end{figure*}

We further investigate the color dependence of the residuals between the observed magnitudes and the standard-star predictions after correcting for $ZP(i)$ and $SS(i,n,nX,nY)$. The results are shown in Figure~\ref{Color_ZP}. We divide all sources into blue and red stars using a boundary of $BP-RP = 1.25$, and compute separate zero-points for blue and red stars in each visit. As seen in the left panel, residual differences between blue and red stars at the level of 1 - 2 per cent persist across visits. With the exception of the first half of the night on 2021-10-1 -- during which the atmospheric conditions were poor and the humidity was high -- the zero-point differences between blue and red stars exhibit a clear correlation with the seeing: when the seeing increases, redder stars appear brighter relative to bluer stars.
A plausible explanation is that the size of the point spread function (PSF) is dominated by atmospheric turbulence, while redder stars have intrinsically smaller PSFs in Subaru/HSC observations (\citealt{HSCPSFs_Carlsten_2018}). Under a fixed aperture, the flux loss therefore becomes less severe for red stars as the seeing increases. Although the $ZP(i)$ and $SS(i,n,nX,nY)$ terms in our calibration model effectively correct for the overall aperture-related flux loss, an additional correction term is required to account for this color-dependent aperture correction.

Given that the typical uncertainty of an individual standard star is 0.04 -- 0.06 mag (see Figure~\ref{SS}), at least $\sim$30 standard stars are required to achieve a correction term with an uncertainty below 1 per cent. Because red stars are far less numerous than blue stars with the current sample and definition of red/blue, we apply the additional correction term $ZP_{\mathrm{Blue/Red}}(i)$ only to those visits in which more than 30 red standard stars are available, as indicated by the results in the left panel of Figure~\ref{Color_ZP}.

\begin{figure*}[htbp!]
\includegraphics[width=180mm]{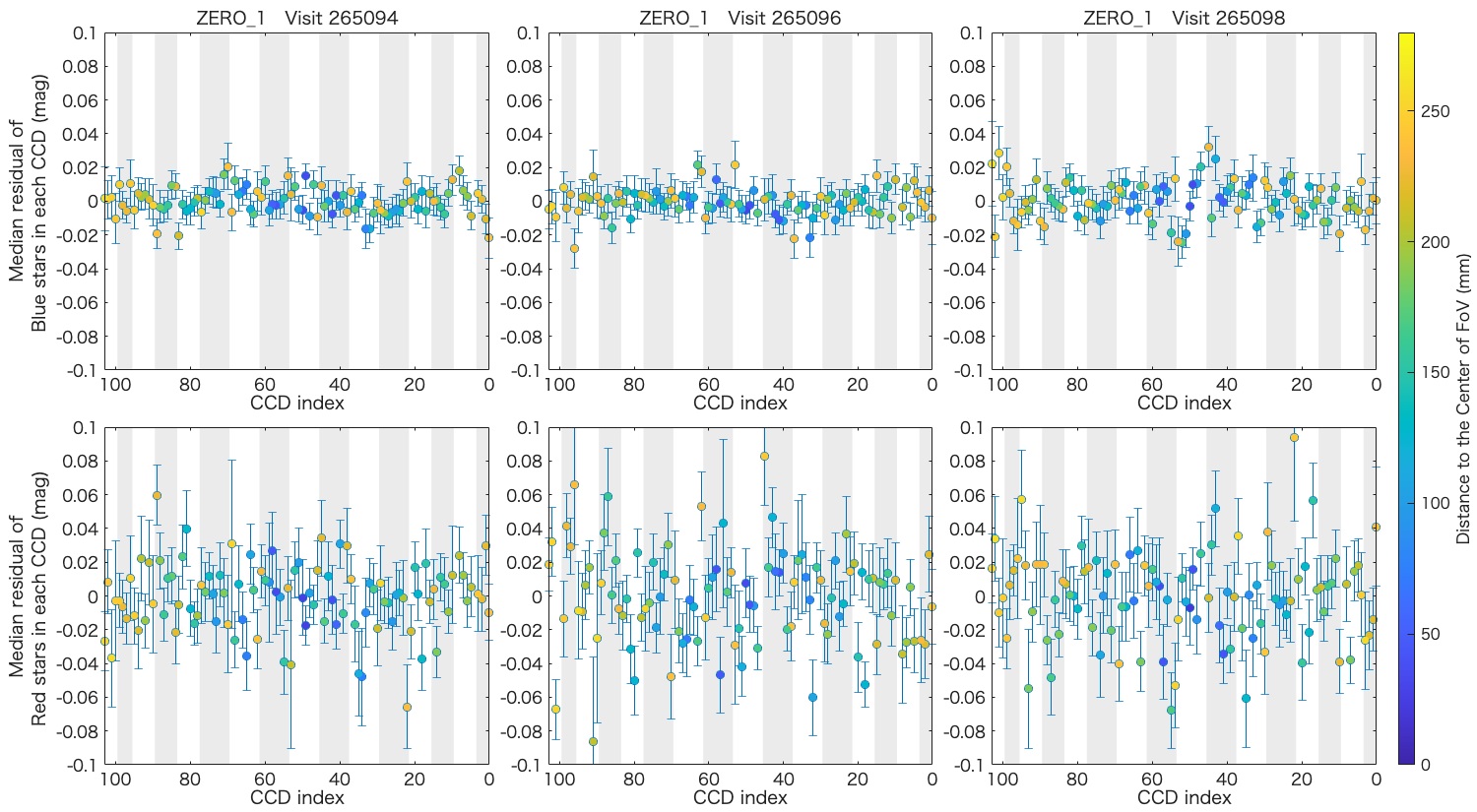}
\caption{Color-dependent zero-points for each CCD as a function of CCD index, shown for three consecutive long-exposure visits of field \texttt{ZERO\_1}. The upper panels correspond to blue stars, and the lower panels to red stars. The color of each point indicates its distance from the FoV center. The error bars represent the uncertainties of the zero-points, estimated from the number of available standard stars.
The alternating gray and white regions denote different CCD columns in the HSC\@. For the ZERO survey observations, the position angle is fixed at $\mathrm{PA}=0^\circ$. Within each column, the CCD index increases from east to west (corresponding to bottom to top in Figure~\ref{SS}), while the columns themselves are ordered from south to north (corresponding to right to left in Figure~\ref{SS}), with the indexing continuing at the bottom of the next column to the left.
For the detailed CCD arrangement, please see the instrument documentation \url{https://www.naoj.org/Instruments/HSC/ccd.html}.}
\label{Color_CCD}
\end{figure*}

In principle, this color-dependent aperture correction should not only affect the zero-points of each visit, but also vary across CCDs within a given visit (i.e., with position across the FoV). However, due to the limited number of available standard stars, we only characterize, rather than correct the color-dependent zero-points (i.e., color-dependent spatial structures) for each CCD. Figure~\ref{Color_CCD} presents the color-dependent spatial structures for three consecutive long-exposure visits of field \texttt{ZERO\_1}, which contains the largest number of standard stars. The typical amplitude of these structures is within the 1 per cent level and shows noticeable temporal variations across the three visits. Manual inspection of all visits confirms that this effect varies with time and is generally continuous across the FoV.

For the field with the highest standard-star density (\texttt{ZERO\_1}), the uncertainties of the CCD-level color-dependent zero-points are comparable to (for blue stars) or even larger than (for red stars) the corrections themselves. This is because the procedure effectively uses standard stars to trace PSF variations across the FoV for each visit, while the intrinsic uncertainties of individual standard stars are larger than the scale of these variations. Consequently, a sufficiently large number of standard stars with reliable color measurements is required to statistically reduce these uncertainties. However, the current number of standard stars allows only a qualitative assessment of the color-dependent spatial structures, and is insufficient for deriving reliable quantitative corrections. Therefore, we refrain from applying further corrections at the CCD level.

\subsection{Result and Comparison}
In the ZERO survey, each source typically has one to four observations. We merge these by computing a weighted mean using the aperture photometric uncertainties as weights, obtaining a combined calibrated magnitude $NB395_{\mathrm{cali\_uni}}$ for each source. However, given the survey's band and depth, the final calibration and photometric precision cannot be directly assessed with standard stars, since both the HSC data and the standard star magnitudes have non-negligible uncertainties.

\begin{figure*}[htbp!]
\includegraphics[width=180mm]{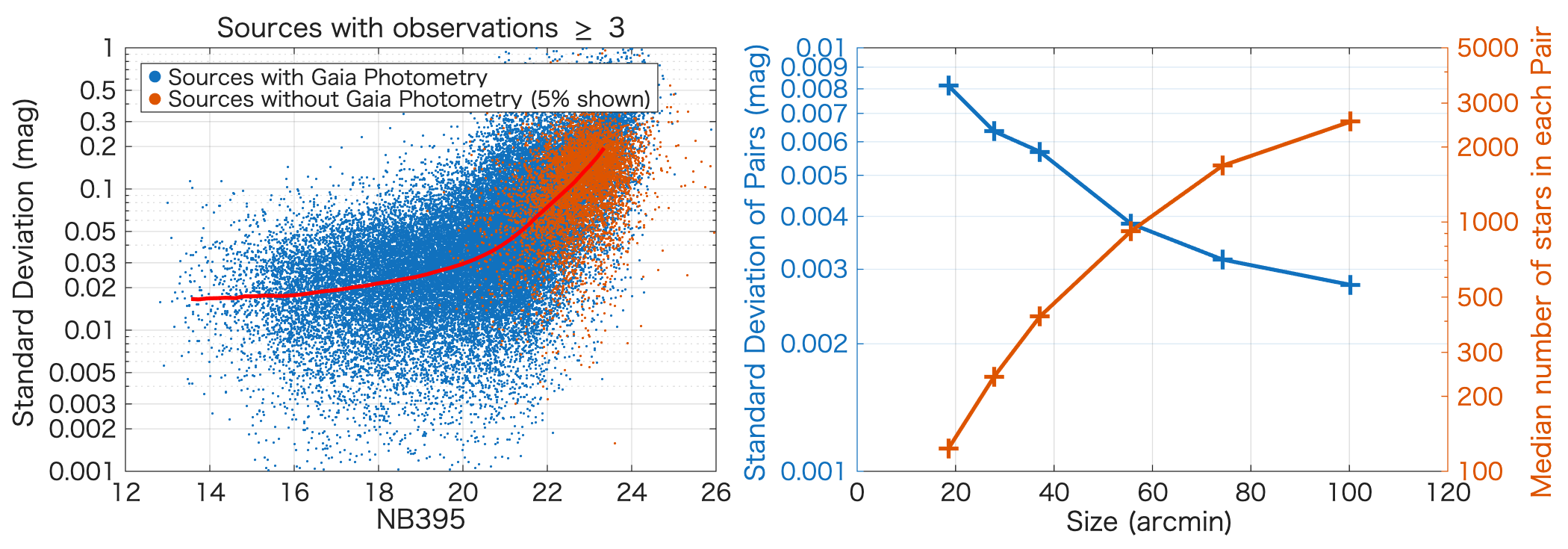}
\caption{The left panel shows the standard deviation of repeated observations as a function of brightness for sources with more than three observations. The right panel presents the medians of the magnitude differences of stars in pairs of visits across different annular spatial bins, along with the typical number of stars in each bin.}
\label{ZERO_precision}
\end{figure*}

The calibration precision is defined as the precision of $ZP(i) + SS(i,n,nX,nY) + ZP_{Blue/Red}(i)$. To remove the effects of measurement uncertainties in both HSC photometry and the standard stars, we assess this precision using two methods based on the internal dispersion of repeated observations.
Firstly, we evaluate the dispersion of repeated observations as a function of $NB395$ for sources with more than three observations, as shown in the left panel of Figure~\ref{ZERO_precision}. At the bright end, where photometric uncertainties are negligible, the standard deviation of the repeated observations asymptotically approaches 0.016 mag, which we adopt as an estimate of the lower limit of the calibration precision. The standard deviation increases to 0.03 mag at $NB395 = 20$ and reaches 0.1 mag at $NB395 = 22.5$, where the photometric uncertainties dominate. It is worth noting that, after combining multiple observations, the uncertainty $NB395_{err}$ is estimated by dividing this standard deviation by $\sqrt{n\_obs}$, where $n\_obs$ is the number of observations.
Secondly, we construct pairs of visits that observed the same fields among the 89 visits, yielding a total of 140 visit pairs. For each pair, we compute the median difference of $NB395_{\mathrm{cali}}$ between the two visits within annular spatial bins of different scales across the FoV, and then evaluate the dispersion among these 70 medians (only for long exposures visits located in the equatorial region) as an indicator of calibration precision. 
Typically, each annular spatial bin contains more than one hundred sources, so the uncertainty of each median is far smaller than that of individual  measurements of a sources. Consequently, the measurement uncertainties of HSC no longer affect the assessment of calibration precision.
Using this method, we derive the dispersion of the median differences at various spatial scales, as shown in the right panel of Figure\,\ref{ZERO_precision}. When the spatial scale approaches the full FoV, the dispersion primarily reflects the precision of $ZP(i) + ZP_{\mathrm{Blue/Red}}(i)$. As the spatial scale decreases, it traces the precision of $ZP(i) + ZP_{\mathrm{Blue/Red}}(i) + SS(i,n,nX,nY)$, with the dispersion gradually increasing from about 0.003 mag to 0.008 mag. It should be noted that this increase not only reflects the variation of calibration precision with spatial scale, but also results from the reduced number of sources per bin. 
Nevertheless, it provides a robust assessment of the relative calibration precision across different spatial scales, demonstrating that our calibration precision is better than 0.01 mag on scales of at least 20 arcmin and larger.

\begin{figure*}[htbp!]
\includegraphics[width=180mm]{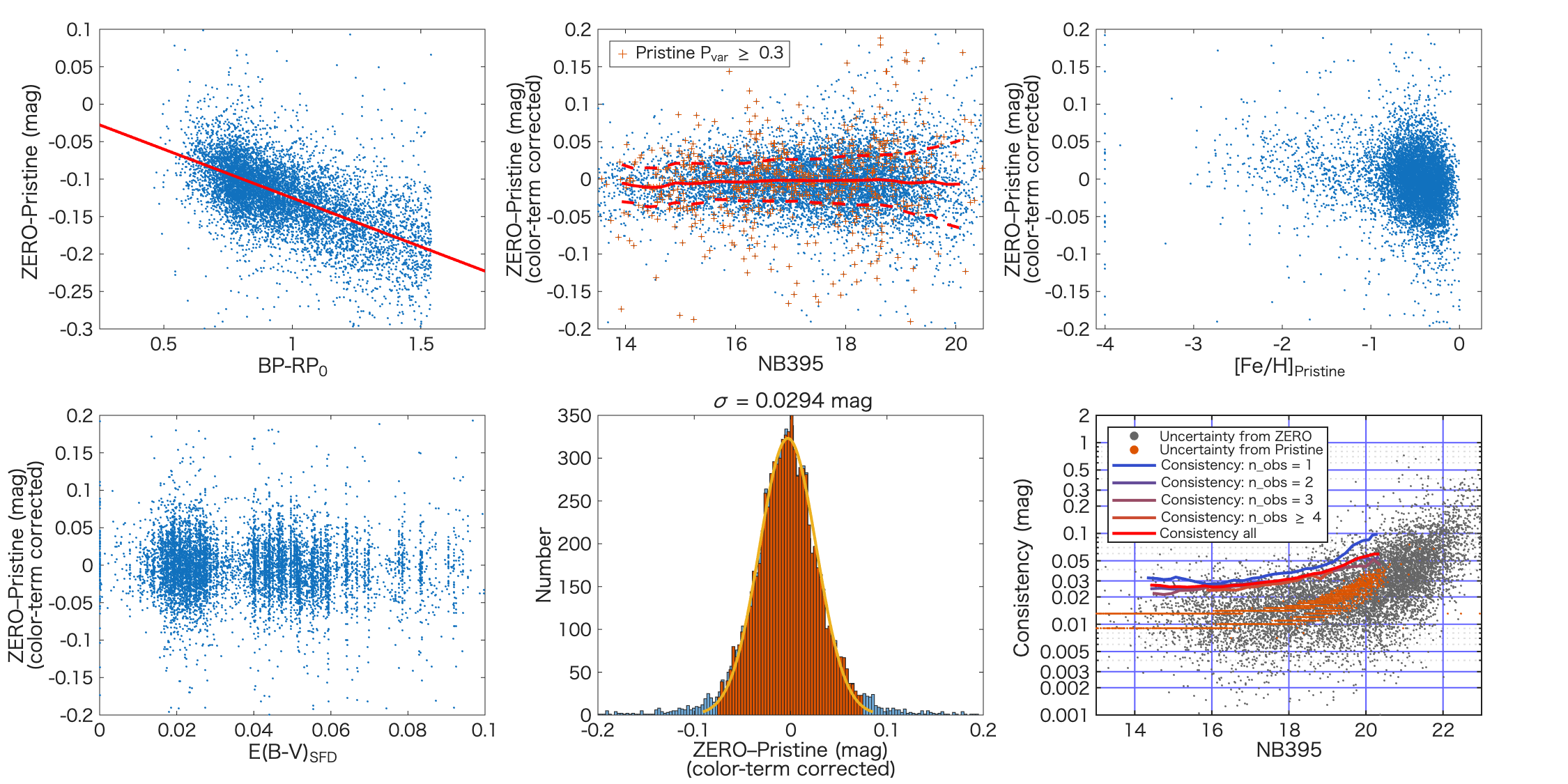}
\caption{The panels show, in sequence, the comparison between the ZERO (NB395 filter) and Pristine photometry: 
the color term and its correction, the magnitude differences as functions of brightness 
(with the red curves indicating the median and the $\pm 1\sigma$ range of the differences, and sources with $P_{\rm var} \geq 0.3$ marked by orange crosses), 
$\mathrm{[Fe/H]}_{\mathrm{Pristine}}$, and $E(B-V)_{\mathrm{SFD}}$, as well as the Gaussian fits to the magnitude differences. 
The bottom-right panel further presents the dispersion between ZERO and Pristine, together with the variation of the uncertainties provided by both surveys as a function of brightness.
}
\label{ZERO_PR}
\end{figure*}

The above assessment provides only a statistical evaluation of the calibration precision on different spatial scales. 
Considering that our $NB395$ filter closely matches the Pristine Ca H\&K filter (\citealt{Pristine_Martin_2024}), a comparison with the Pristine survey serves as a good external dispersion check. However, the two filters are not exactly identical, and a color term exists between them. In addition, the Pristine catalog provides intrinsic magnitudes. Therefore, we apply an extinction correction using the two-dimensional dust map of SFD98 together with an empirical constant extinction coefficient of 4.5, calibrated through comparison with the Pristine survey and broadly consistent with expectations from standard extinction curves for the NB395 band, and further apply a color-term correction on our $NB395$ photometry. The comparison, shown in Figure~\ref{ZERO_PR}, demonstrates excellent agreement, with a scatter of only 0.029 mag between our calibrated $NB395_{\mathrm{cali\_uni}}$ photometry and the Pristine measurements.
The bottom-right most panel of Figure~\ref{ZERO_PR} shows the dispersion between our ZERO and the Pristine photometry as a function of brightness. 
We evaluate the dispersion between our ZERO and the Pristine photometry for sub-samples with different brightness levels and different numbers of observations (\texttt{n\_obs = }1 -- 4, and averaged), as illustrated by the curves from blue to red in the figure.
When only a single observation is available in ZERO, the precision of our $NB395$ photometry is slightly reduced. 
With multiple observations, the agreement is better than 0.03 mag at $NB395 < 18$ and better than 0.05 mag at $NB395 < 20$. We also show the uncertainty estimates provided by our calibration and those from Pristine, which exhibit comparable results.

The Pristine catalog also provides $P_{\rm var}$, which represents the probability that a source is variable based on the multi-epoch \textit{Gaia} photometry (\citealt{Pristine_Martin_2024}). Following their recommended criterion, we highlight sources with $P_{\rm var} \geq 0.3$ in the top-middle panel of Figure~\ref{ZERO_PR}. Although the overall increase in the ZERO--Pristine dispersion toward fainter magnitudes is primarily driven by the lower photometric signal-to-noise ratio, stellar variability can explain some of the outliers at the bright end. In particular, most of the small number of extreme outliers have high $P_{\rm var}$, suggesting that these extreme discrepancies are likely caused by stellar variability, including the largest difference of approximately 0.83 mag.

\subsection{Extended Sources}
We set a \texttt{Galaxy\_candidate} flag to indicate whether a source is extended. 
For sources with available Gaia photometry, this flag is assigned based on $\texttt{phot\_bp\_rp\_excess\_factor} > 2$. 
For sources without Gaia photometry, the flag is determined using the ZERO survey data by comparing the flux measured within a larger aperture to that measured within the fixed 12-pixel (2 arcsec) aperture; the flag is set when the larger-aperture flux is brighter by more than 0.3 mag. This provides a supplementary classification for sources without Gaia photometry, as neither Gaia nor HSC-SSP broadband photometry is available for all ZERO sources.

Based on this criterion, the point/extended source ratio decreases with magnitude: it falls below unity at $NB395 \sim 22$ and below 0.5 at $NB395 \sim 23$, indicating that the majority of faint sources are classified as extended. 
Random visual inspections using HSC DR2 Wide layer $g$ band images confirm that most extended sources are galaxies. 
However, this flag becomes increasingly affected by selection effects at faint magnitudes, where larger photometric uncertainties tend to cause sources to be misclassified as extended. 
In subsequent analyses of metallicity measurements, additional stellar selection criteria based on higher signal-to-noise broadband photometry from the HSC-SSP will therefore be applied.
In addition, for extended sources, the fixed aperture adopted in this work inevitably misses a substantial fraction of the total flux. As the primary focus of this study is the metallicity of stars, we do not attempt to further correct the photometry of extended sources and only provide a preliminary point/extended classification.

\section{[Ca/H] Measurement}
In this section, we describe our measurement of [Ca/H] using the ZERO survey $NB395$ band together with the $g$ and $i$ bands from the HSC-SSP Wide survey and XPSP, combined with the a new technique we have developed and call the "Physics-constrained Stellar Locus Fitting Method". 

\begin{figure}[htbp!]
\includegraphics[width=90mm]{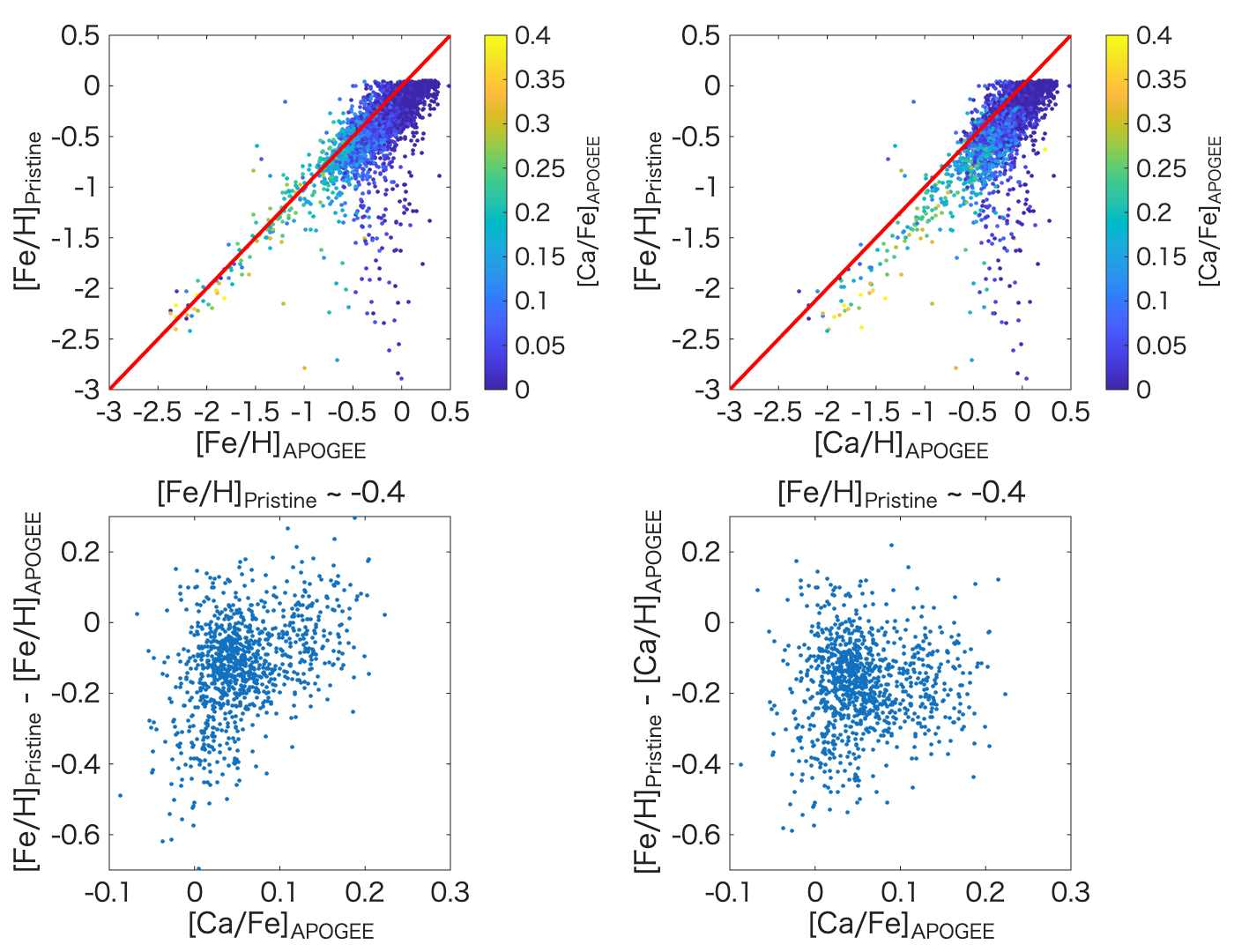}
\caption{The panels show the comparisons of \([{\rm Fe/H}]_{\rm Pristine}\) with \([{\rm Fe/H}]_{\rm APOGEE}\) and \([{\rm Ca/H}]_{\rm APOGEE}\). The plotted points represent the common sources between Pristine and APOGEE, selected to satisfy \({\rm SNR} > 400\) and \(\log g < 3\) in APOGEE.}
\label{Metal_PR_APOGEE_cmp}
\end{figure}

In previous photometric metallicity studies (e.g., \citealt{GALEX_Lu_2024,Pristine_Martin_2024}), when only a single metal-sensitive band (e.g., $NUV$) is employed, the number of measurable parameters is limited to one. Because iron absorption lines are widely distributed across the near-ultraviolet and optical bands, the metallicity derived in such cases is typically [M/H] or [Fe/H]\@. Only through the use of multiple photometric bands can the abundances of additional elements such as C, N, O, Mg, and Ca be constrained (e.g. \citealt{JPLUS_Yang_2022}).
Even in the Pristine survey, which uses a narrow-band filter targeting the Ca H\&K absorption lines, the primary observable is the Ca H\&K feature, yet the final reported quantity is still an estimate of [Fe/H]\@. The Ca H\&K lines are among the strongest metal absorption features in the optical spectra of metal-poor stars and have therefore long been used as metallicity tracers in low-resolution spectroscopy and narrow-band photometric surveys. This practice implicitly assumes that [Ca/Fe] is constant, allowing the directly measured [Ca/H] to be converted into [Fe/H] through a simple transformation. While this assumption is generally valid in the metal-rich galactic disk or the metal-poor halo, [Ca/Fe] is not constant over a wider [Fe/H] range. 
Consequently, the inferred [Fe/H] may exhibit systematic biases driven by variations in [Ca/Fe]. To be fair, the broadened Ca H\&K feature depends not only on the Ca abundance itself, but also on the continuous opacity, which is sensitive to the overall metallicity. Nevertheless, the Ca H\&K absorption remains the dominant metal-sensitive feature probed by the Pristine filter.
For example, in Figure~\ref{Metal_PR_APOGEE_cmp}, we present a comparison of the metallicities for approximately 1,000 giants common sources between Pristine DR1 and APOGEE (\citealt{APOGEE_Majewski_2017}). The results show that the [Fe/H] provided by Pristine are more consistent with APOGEE [Ca/H] -- apart from a constant offset -- than with APOGEE [Fe/H]\@. Moreover, at \([{\rm Fe/H}] \sim -0.4\), the quantity \([{\rm Fe/H}]_{\rm Pristine} - [{\rm Fe/H}]_{\rm APOGEE}\) exhibits a clear correlation with [Ca/Fe]. Nevertheless, at the low metallicities most relevant to the Pristine survey goals, the Pristine [Fe/H] estimates remain in good agreement with APOGEE [Fe/H], although [Ca/Fe] is not strictly constant even within the halo population, since Ca and Fe originate from different nucleosynthetic channels and enrichment timescales.

Given the similarity between the ZERO survey $NB395$ band and the Pristine Ca H\&K narrow-band filter, we adopt [Ca/H] as the primary metallicity indicator in the following analysis, since it provides a more direct measurement that is less dependent on assumptions about [Ca/Fe].
A model-based investigation into the chemical abundance indicator directly probed by the $NB395$ band is presented in the Appendix.

\subsection{Data Selection}
Because the stellar locus method requires at least one color to characterize stellar effective temperature, and the ZERO survey field lies within the HSC-SSP footprint where the magnitude range covers fainter stars well, we use the HSC-SSP Wide-layer $g-i$ color as our temperature indicator.
However, some regions lie outside the HSC-SSP footprint, and HSC-SSP photometry saturates at $NB395 < 18$. For these sources, we therefore supplement the temperature indicator by synthesizing the $g-i$ color with the XPSP method, using Gaia XP spectra and the HSC-SSP transmission curves. We then apply the following selection criteria to ensure the sources are stars and that their $NB395$, $g$, and $i$ photometry is reliable:

\begin{enumerate}[itemsep=0.2cm, topsep=0.5cm, parsep=0.2cm, partopsep=0.3cm]
    \item For sources with Gaia counterparts, we require $\texttt{phot\_bp\_rp\_excess\_factor} \leq 2$. For sources with HSC-SSP Wide-layer broadband photometry, we require $\texttt{i\_extendedness\_flag}=0$ and $\texttt{prob\_star} > 0.9$ from the HSC photo-$z$ group (\citealt{HSCPhotoz}) in order to select stellar sources.
    \item For the HSC-SSP photometry, we employ the three color - color diagrams, $g-i$ vs.\ $g-r$, $g-i$ vs.\ $g-z$, and $g-i$ vs.\ $g-y$, to exclude sources that show significant deviations from the stellar locus, thereby removing objects with unreliable HSC-SSP photometry. 
    \item For the XPSP $g$ and $i$ photometry, sources with large $\texttt{phot\_bp\_rp\_excess\_factor}$ values or those significantly deviating from the general relation between $\texttt{phot\_bp\_rp\_excess\_factor}$ and the $g-i$ color are removed. 
    Specifically, sources satisfying
    \begin{equation*}
    \begin{aligned}
        \texttt{phot\_bp\_rp\_excess\_factor} > \\
        0.0067\times(g-i)^3 - 0.0049\times(g-i)^2 \\
        - 0.0508\times(g-i)+1.1720
    \end{aligned}
    \end{equation*}
    are excluded.
\end{enumerate}

\begin{figure}[htbp!]
\includegraphics[width=90mm]{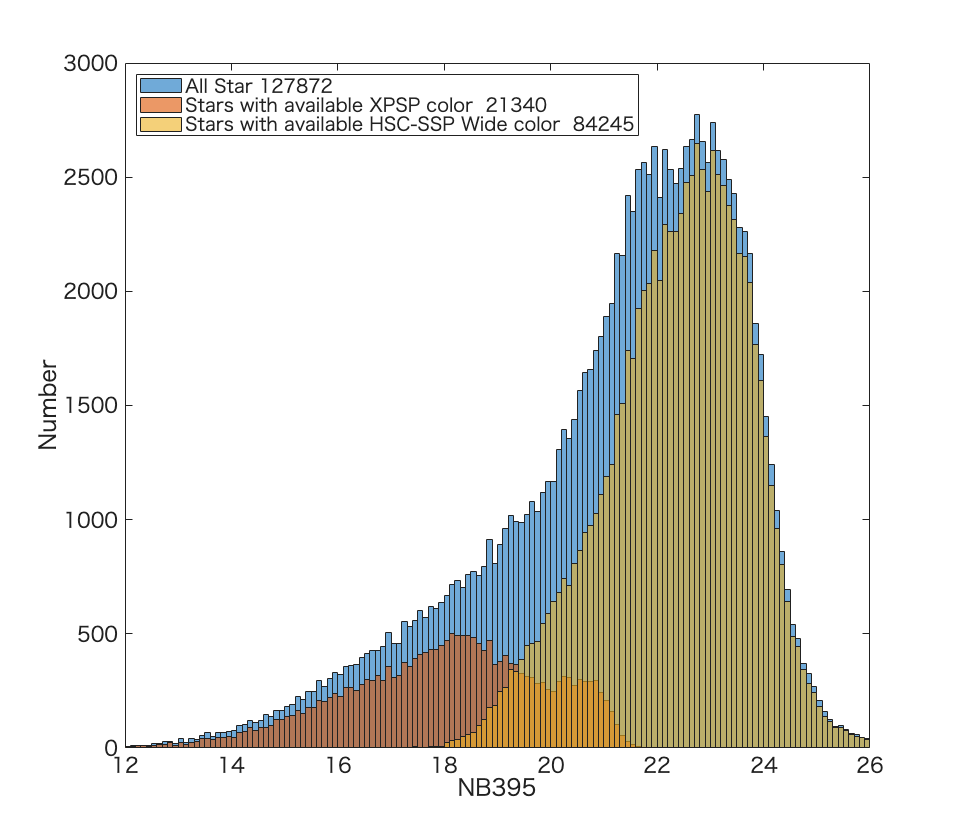}
\caption{The magnitude distribution of stars in the ZERO survey in the NB395 band.}
\label{Metal_number}
\end{figure}

Based on the above selection criteria, we obtain a sample of 127,872 stars with Gaia XP or HSC-SSP Wide-layer counterparts. Among them, 21,340 sources have reliable $g-i$ colors from XPSP, and 84,254 from the HSC-SSP observations, as shown in Figure~\ref{Metal_number}. For sources with both measurements available, we retain only the HSC-SSP values, as the Gaia XP uncertainties are generally larger for these faint stars. For the $g-i$ colors synthesized from Gaia XP, we apply a color correction based on the small number of sources with both Gaia XP and HSC-SSP measurements to ensure consistency with the HSC-SSP photometric system.

For the extinction correction of the $NB395$, $g$, and $i$ bands, we adopt the SFD98 two-dimensional dust map, which provides reliable extinction estimates given that the ZERO survey region lies at high Galactic latitude with typically low extinction ($E(B-V) \sim 0.02$ mag).
For the extinction coefficients, we use a fixed value of 4.5 for the $NB395$ band, while for the $g$ and $i$ bands we adopt the temperature- and extinction-dependent coefficients provided by \cite{ExtCoeff_Zhang_2024}.

\begin{figure}[htbp!]
\includegraphics[width=90mm]{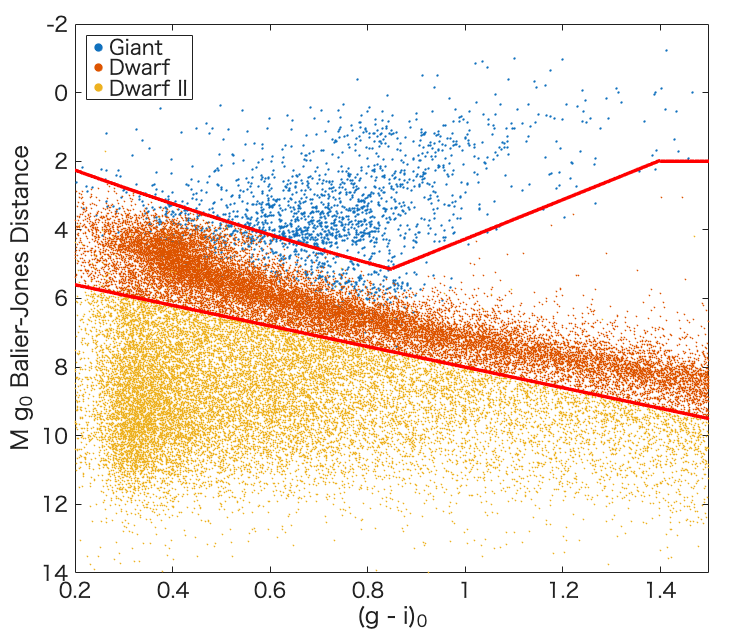}
\caption{The Hertzsprung-Russell diagram of stars in the ZERO survey. The red curves delineate the boundaries between the different regions.}
\label{HRD}
\end{figure}

The Hertzsprung -- Russell Diagram (HRD) of the current stellar sample is presented in Figure~\ref{HRD}, where the absolute magnitude $M g_0$ is derived using the distances provided by \cite{Distance_BJ_2021}.
Based on the red curves shown in the figure, and within the range $0.2 < (g-i)_0 < 1.5$, we divide the sample into several subsamples, excluding M dwarfs whose colors may be affected by molecular-band variations. Since $M g_0$ is significantly influenced by distance uncertainties, a star is classified as a giant whenever the upper bound of its $M g_0$ uncertainty lies above the upper red curve, while the remaining sources are considered dwarfs.
Sources located below the lower red line and with $(NB395 - g)_0 < 0.3$, corresponding to ultraviolet-bright objects, are assigned to the Dwarf II region in order to minimize white dwarf \textbf{and quasar} contamination in the main Dwarf region. Compared to the Dwarf region, the Dwarf II region is therefore expected to contain a higher fraction of white dwarfs.
Sources with Gaia counterparts but without Gaia distance measurements are assigned to the No Gaia Distance sample, while sources without Gaia counterparts are classified into the No Gaia Photometry sample.

\begin{figure*}[htbp!]
\includegraphics[width=180mm]{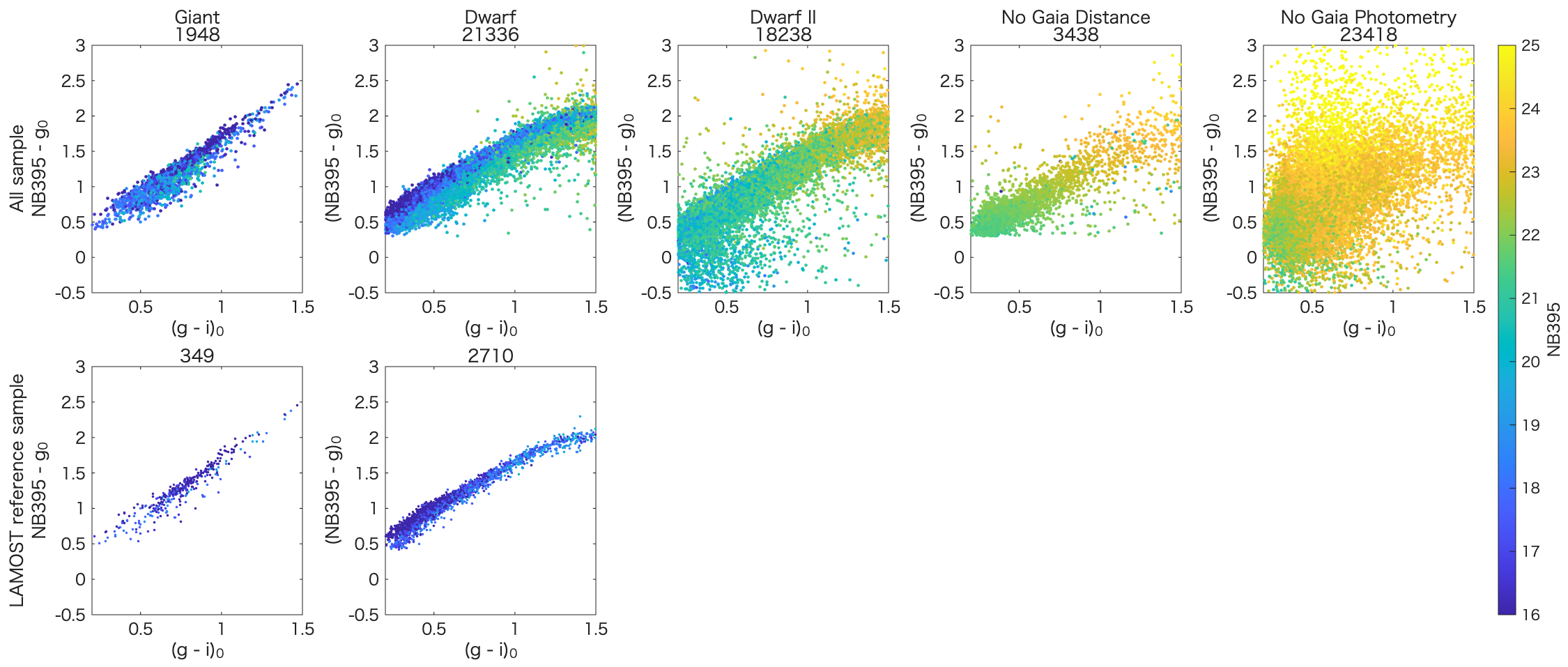}
\caption{Color-color diagrams for five samples: Giants, Dwarfs, Dwarf II, and Distance-unavailable sources.
The top row shows all sources in each sample, while the bottom row presents the corresponding reference samples, consisting of sources cross-matched with LAMOST\@. The color of the points indicates the $NB395$ magnitude. The number of sources in each sample is labeled at the top of each panel.}
\label{Metal_sample}
\end{figure*}

In total, the final sample for which we derive measurements of [Ca/H] consists of 1,948 giants, 21,336 dwarfs, 18,238 Dwarf II, 3,438 No Gaia Distance, and 23,418 No Gaia Photometry sources. Their color-color diagrams are shown in Figure~\ref{Metal_sample}.
As shown in Figure~\ref{Metal_sample}, sources that can be classified as giants or dwarfs are generally bright and have high signal-to-noise ratios, forming clearly separated stellar loci, particularly at the red end ($(g-i)_0 > 1$). 
For the Dwarf II sample, the sources are fainter, resulting in larger uncertainties. Consequently, some outliers deviate from the stellar locus, and a substantial number of sources with $(NB395 - g)_0 < 0.3$ are consistent with the location expected for white dwarfs in color–color space.
For the faint sources in the No Gaia Photometry sample, the dispersion is noticeably larger. However, this dispersion is mainly dominated by sources with only a single observation (exposure time $<20$ minutes) at faint magnitudes ($NB395 > 23$), which account for approximately 34 per cent of this sample. For sources with longer exposures or brighter magnitudes, the dispersion is comparable to, or only slightly larger than, that of the No Gaia Distance sample.

We further cross-match our sample with LAMOST DR9 and select common sources with $S/N > 10$ as the reference sample for subsequent stellar locus fitting. The color-color diagrams of the giant and dwarf sources in the reference sample are shown in the bottom row of Figure~\ref{Metal_sample}.

\subsection{Physics-constrained Stellar Locus Fitting Method}
We employ a Physics-constrained Stellar Locus Fitting Method to construct empirical stellar locus models that map stellar colors to metallicity for giants and dwarfs separately:
\begin{equation}
\begin{aligned}
(NB395-g)_0   & = \mathcal{F}((g-i)_0,{\rm [Ca/H]})
\end{aligned}
\label{EQ3}
\end{equation}
where $\mathcal{F}$ denotes a fourth-order bivariate polynomial. 
For the dwarf sample, owing to its broader color range, the stellar locus exhibits a more complex, higher-order structure (\citealt{Metal_Huang_2025}). As a result, the metallicity-independent relation between $(NB395 - g)_0$ and $(g - i)_0$ in the function $\mathcal{F}$ is derived through numerical interpolation rather than an explicit analytic form. 

The stellar locus method relies on obtaining strong constraints on the coefficients of $\mathcal{F}$. Its main limitation is that reference samples based on spectroscopic stellar parameters mostly cover the metal-rich regime, while reference stars in the crucial metal-poor regime (e.g., [Fe/H] $<-2$ or [Fe/H] $<-3$) are extremely rare. Consequently, the stellar locus is poorly constrained there.

The Physics-constrained Stellar Locus Fitting Method tackles this by constraining the stellar locus with both spectroscopic atmospheric parameters and stellar atmosphere models.
This approach is conceptually similar to the use of theoretical spectral gradients as physical constraints in data-driven spectroscopic stellar-label determination (e.g., \citealt{Ting2017b,Xiang2019}). Here, we use the stellar atmosphere models to constrain the response of photometric colors to metallicity rather than directly adopting their absolute synthetic colors.
We primarily rely on the common sources between our sample and the DD-Payne catalog for LAMOST DR9, adopting their spectroscopic [Ca/H] values as reference. 

In addition, we compute $(g-i)_0$ and $(NB395-g)_0$ using \texttt{turbospectrum} (\citealt{Turbospectrum_Plez_2012}) with the Vienna Atomic Line Data Base (VALD; \citealt{VALD_Piskunov_1995}) and MARCS model atmospheres (\citealt{MARCS_Gustafsson_2008}), and evaluate 
$\frac{\partial (NB395-g)_0}{\partial \mathrm{[Ca/H]}}$
at each value of $(g-i)_0$. 

We require that, for $\mathrm{[Ca/H]} < -3$, 
$\frac{\partial \mathcal{F}}{\partial \mathrm{[Ca/H]}}$
be consistent with the model-predicted 
$\frac{\partial (NB395-g)_0}{\partial \mathrm{[Ca/H]}}$
thereby providing an effective constraint on the stellar locus model in the metal-poor regime.
Specifically, we analytically differentiate the polynomial $\mathcal{F}$ with respect to [Ca/H] and evaluate the resulting derivative at the $(g-i)_0$ and [Ca/H] values of the selected stellar atmosphere models. These derivative constraints are then fitted simultaneously with the observed stellar colors to determine the polynomial coefficients. Thus, the spectroscopic reference sample primarily determines the empirical stellar locus, while the stellar atmosphere models constrain its metallicity dependence in the metal-poor regime. This is only one possible way to incorporate the physical constraints; alternatively, the model-predicted derivatives could be imposed as explicit constraints in a constrained optimization framework.

Considering that $\log g$ has a significant impact on the $(NB395-g)_0$ color, and that extremely metal-poor stars exhibit a clear correlation between $\log g$ and color, we compute isochrones of 12 Gyr for metallicities [M/H] = $-$2 and [M/H] = $-$1.4 using the default solar-scaled PARSEC models (\citealt{PARSEC_Bressan_2012,PARSEC_Nguyen_2022,PARSEC_Nguyen_2025}). The MARCS model grid points are selected only within the region bounded by these two isochrones.

\subsection{Result}
\begin{figure*}[htbp!]
\includegraphics[width=180mm]{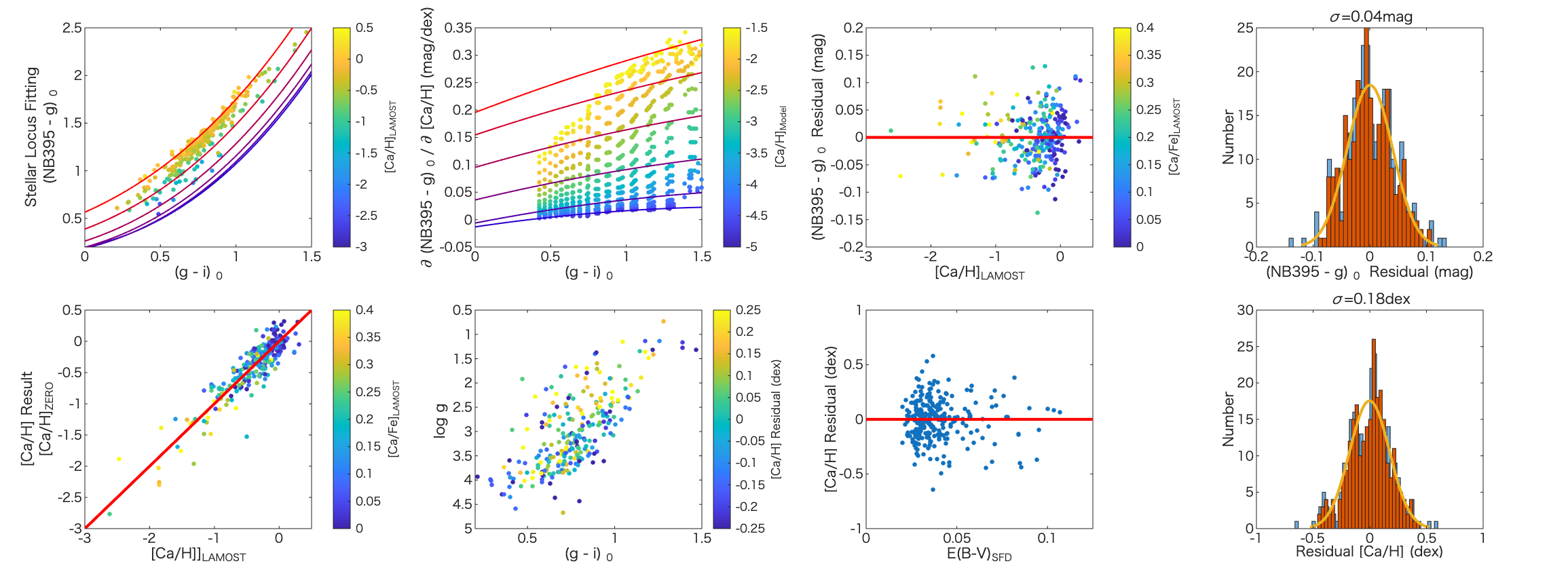}
\caption{Results of the stellar locus fitting and the derived [Ca/H] estimates for Giant Model.
The top row shows the results of the stellar locus fitting, including the fitted stellar loci and their partial derivatives $\partial(NB395-g)_0/\partial[\mathrm{Ca/H}]$. In the first two panels, the curves from red to blue correspond to [Ca/H] = 0 to $-5$ with intervals of 1 dex, respectively. The third panel shows the distribution of the $(NB395 - g)_0$ residuals as a function of $[{\rm Ca/H}]_{\rm LAMOST}$, and the fourth panel presents the histogram of the residuals.
The bottom row presents a comparison between the photometrically estimated [Ca/H] and the LAMOST measurements. From left to right, the panels show the direct comparison between the estimated [Ca/H] and LAMOST [Ca/H], the distribution of the [Ca/H] residuals across the HR diagram, as a function of $E(B-V)$, and the corresponding histogram. The dispersion in each distribution is labeled at the top of each panel and is computed from the red bars after applying a $3\sigma$ clipping.}
\label{Fitting_Giant}
\end{figure*}

\begin{figure*}[htbp!]
\includegraphics[width=180mm]{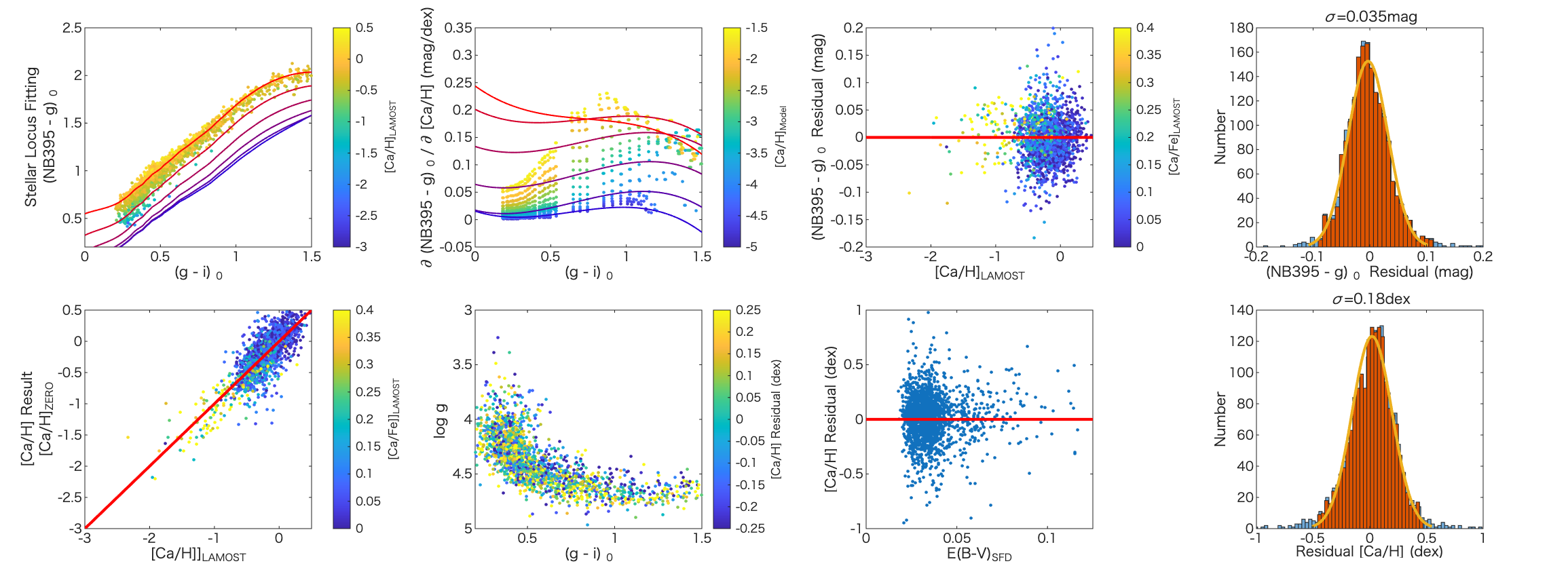}
\caption{Same as Figure \ref{Fitting_Giant}, but for dwarf sample.}
\label{Fitting_Dwarf}
\end{figure*}

For the giant and dwarf samples, the stellar locus models, their partial derivatives, and fitting residuals are shown in Figures~\ref{Fitting_Giant} and~\ref{Fitting_Dwarf}, respectively. The stellar locus model follows the metal-rich LAMOST reference sample and agrees with the MARCS predictions at [Ca/H] $< -3$.

The stellar locus residuals show no systematics with [Ca/H] or [Ca/Fe]. The dispersion of the $(NB395 - g)_0$ color residuals is 0.040 mag for giants and 0.035 mag for dwarfs. Our [Ca/H] values agree well with $[{\rm Ca/H}]_{\rm LAMOST}$, with an overall dispersion of 0.18 dex for both giants and dwarfs. These dispersions and [Ca/H] consistencies include LAMOST measurement uncertainties and systematic biases from $\log g$ variations. As shown in the HRDs in Figures~\ref{Fitting_Giant} and~\ref{Fitting_Dwarf}, our [Ca/H] estimates show systematic trends with $\log g$, especially for giant stars. Red clump (RC) stars, whose locus differs from that of typical RGB stars, also display systematic [Ca/H] offsets (\citealt{Metal_Huang_2025}), as seen in Figure~\ref{Fitting_Giant}.
Because current metallicity estimates use only one metal-sensitive band and Gaia distances are too uncertain for reliable luminosity constraints, we do not attempt additional $\log g$ systematics corrections. Instead, we use LAMOST $\log g$ for giants only as auxiliary information to gauge the effect of surface gravity on dispersion. Adding LAMOST $\log g$ to the stellar locus fitting improves agreement with LAMOST [Ca/H] to 0.13 dex.

Furthermore, using the photometric uncertainties in $NB395$, $g$, and $i$ provided by the ZERO survey, XPSP, and HSC-SSP, we estimate the 16th and 84th percentile confidence bounds of estimated [Ca/H] through Monte Carlo simulations. 
The uncertainty in [Ca/H] depends jointly on the photometric uncertainties in the three surveys and on the sensitivity of the color -- metallicity relation at different colors and metallicities. Owing to the strong variation in this sensitivity, the inferred [Ca/H] uncertainty distributions can become highly asymmetric, particularly in the extremely metal-poor regime and at faint magnitudes, making the uncertainty estimation non-trivial. Using the HSC-SSP broad-band photometry and requiring at least three ZERO observations for sources with $(g-i)_0 < 0.8$, the typical [Ca/H] uncertainty is approximately 0.08 dex at [Ca/H] $=-1$, 0.13 dex at [Ca/H] $=-2$, and 0.2 dex at [Ca/H] $=-3$ for sources with $NB395 = 20$. For fainter sources with $NB395 = 23$, the corresponding uncertainties increase to approximately 0.4 dex at [Ca/H] $=-1$ and 0.7 dex at [Ca/H] $=-2$.

\begin{figure*}[htbp!]
\includegraphics[width=180mm]{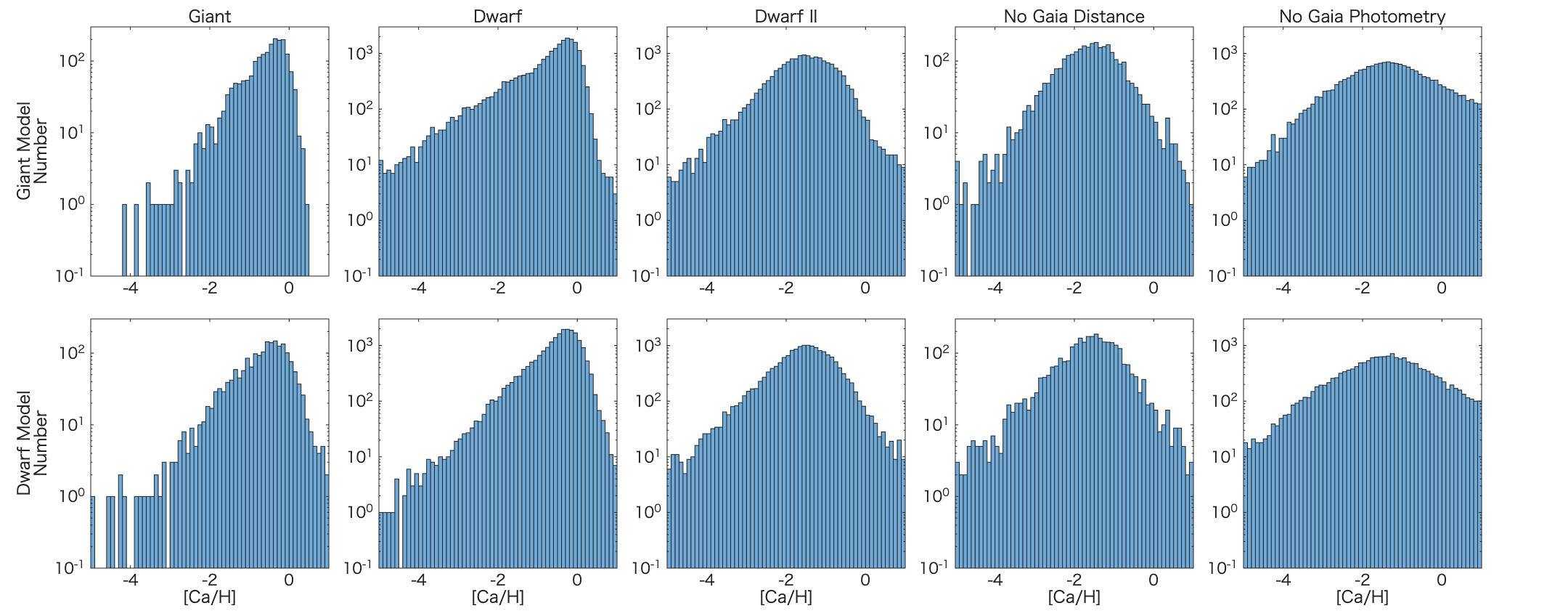}
\caption{The estimated [Ca/H] distributions for five samples: Giants, Dwarfs, Dwarf II, No Gaia Distance, and No Gaia Photometry sources.}
\label{Metal_histo}
\end{figure*}

Figure~\ref{Metal_histo} shows the [Ca/H] distribution for the final sample. For each object, we obtain metallicity estimates from both the dwarf and giant models. When reliable distances are available, the model should be chosen according to the distance-based dwarf/giant classification. Without reliable distances, separating dwarfs from giants is difficult because their loci diverge mainly at the red end. For colors near the main-sequence turn-off ($0.4 < (g-i)_0 < 0.6$), the dwarf and giant metallicity estimates are similar, so the giant-model result may be preferred. If the giant-model estimate lies outside the valid range ([Ca/H] $<-5$ or [Ca/H] $>+1$) while the dwarf-model estimate is within it, the dwarf-model result should instead be adopted.

As shown in Figure~\ref{Metal_histo}, the giant and dwarf samples peak at ${\rm [Ca/H]} \sim -0.3$ and decline toward the metal-poor end with an approximately power-law-like behavior.
In contrast, the Dwarf II, No Gaia Distance, and No Gaia Photometry samples peak at ${\rm [Ca/H]} \sim -1.5$, mainly due to the selection criteria: these generally fainter, and thus statistically more distant, sources are more likely to be halo stars, which typically have lower metallicities.
The distribution also broadens as photometric uncertainties increase, pushing some sources beyond the measurement boundaries; this is an observational effect caused by larger metallicity errors.

\section{Summary}
This study introduces the ZERO survey, a narrow-band photometric program specifically designed to infer stellar metallicities from calcium-sensitive narrow-band imaging conducted along the celestial equator. The survey utilizes the dedicated NB395 filter on the Hyper Suprime-Cam instrument mounted on the Subaru Telescope and integrates these observations with broad-band photometry from the Gaia XP Spectrophotometer (XPSP) and the HSC-SSP Wide layer. This combined dataset enables the efficient detection and detailed characterization of metal-poor stellar populations.

The current ZERO data set consists of 21 HSC pointings covering approximately 35 deg$^2$ along the celestial equator. Owing to the large aperture of the Subaru Telescope and the wide field of view of HSC, the survey reaches depths of NB395 $\sim$ 24.5 mag at the 10$\sigma$ level in one hour of integration, allowing metal-poor stars with typical colors to be detected down to $G \sim 22.5$-$23.4$ mag and substantially extending the distance range accessible to photometric metallicity studies beyond that of existing wide-field surveys. At these depths, red-giant stars are detectable out to $\sim1$~Mpc, while dwarf stars can be detected to $\sim100$~kpc.

We describe the photometric calibration of the NB395 band based on Gaia XP spectra and the recalibrated SDSS Stripe 82 standard star catalog. The resulting calibration achieves a typical zero-point precision of approximately 0.3 per cent, with flat-field precision better than 1 per cent on spatial scales larger than 8 arcmin. Under typical observing conditions with three repeated observations, the photometric precision reaches approximately 1.1 per cent at NB395 $\approx$ 12, 1.5 per cent at NB395 $\approx$ 20, and about 10 per cent at NB395 $\approx$ 24.

We further develop a Physics-constrained Stellar Locus Fitting Method to estimate stellar calcium abundances using NB395 and broad-band colors. The method extends the classical stellar locus approach by introducing model-based constraints, which stabilize the empirical stellar locus in the metal-poor regime.
The resulting photometric [Ca/H] estimates achieve a precision better than 0.18 dex, as measured by the dispersion relative to spectroscopic measurements from LAMOST, for sources brighter than NB395 $\sim$ 20 and span a wide metallicity range from metal-rich stars to the very metal-poor regime ($-3 \le$ [Ca/H]$\le +0.5$). The analysis demonstrates that the NB395 band provides a more direct tracer of [Ca/H] than of [Fe/H].

Based on these methods, we have obtained  calibrated NB395 measurements for approximately one million sources observed over four nights, and a photometric metallicity catalog comprising 61,845 sources with reliable [Ca/H] estimates. 
Because the photometric separation between faint giants and dwarfs is intrinsically ambiguous, photometric metallicity estimates are derived from both giant and dwarf stellar locus models for each source to facilitate flexible scientific applications. 
We plan to make the ZERO survey data, including the source catalog presented in this work, publicly available through a dedicated ZERO survey website (\url{https://sites.google.com/view/zerosurvey/home}), which is currently under preparation. The data will be released as early as possible, although a specific release date has not yet been determined. The metallicity-fitting code used in this work will also be made available through the ZERO survey website. In addition, the code will also be publicly released at \url{https://doi.org/10.5281/zenodo.24079010}.

The ZERO survey provides a new, well-calibrated photometric data set for metallicity studies of distant stellar populations. 
Future extensions of the survey, including observations of additional fields along the HSC-SSP Wide Layer to provide broader sky coverage for studies of the Galactic halo and the incorporation of the NB430 filter targeting the CH absorption feature to provide additional sensitivity to carbon abundance, will further enhance its capability to identify chemically peculiar stars and to investigate the structure and chemical evolution of the Galactic halo.

\vfill\eject
\appendix
\setcounter{secnumdepth}{0}
\setcounter{table}{0}   
\setcounter{figure}{0}
\renewcommand{\thetable}{A\arabic{table}}
\renewcommand{\thefigure}{A\arabic{figure}}

\section{A. The properties of NB395}
As discussed in Section 5, narrow-band photometry targeting the Ca H\&K lines is better suited to trace [Ca/H] rather than [Fe/H]\@. However, the NB395 filter also covers absorption lines of other elements.
In addition, even the Ca H\&K feature itself is sensitive not only to the Ca abundance, but also to the overall metallicity through the continuous opacity and damping terms. 
As a result, the NB395 photometry, or equivalently the $(NB395-g)_0$ color, likely reflects a combination of Ca and overall metallicity effects, with Ca being dominant. In this section, we use MARCS models to perform a preliminary investigation of the chemical abundance information directly encoded in the $(NB395-g)_0$ color.
The following abundance pattern are adopted in the synthetic calculations: the $\alpha$-elements (O, Ne, Mg, Si, S, Ar, Ti) follow $[\alpha/\mathrm{Fe}] = +0.4$ for $\mathrm{[Fe/H]} < -1$, 
$[\alpha/\mathrm{Fe}] = -0.4 \times \,\mathrm{[Fe/H]}$ for $-1 < \mathrm{[Fe/H]} < 0$, and $[\alpha/\mathrm{Fe}] = 0$ for $\mathrm{[Fe/H]} > 0$, while all other elements are assumed to be  $[\mathrm{X/Fe}] = 0$. This pattern approximately follows the typical chemical trends observed in Milky Way stellar populations.

\begin{figure}[htbp!]
\includegraphics[width=90mm]{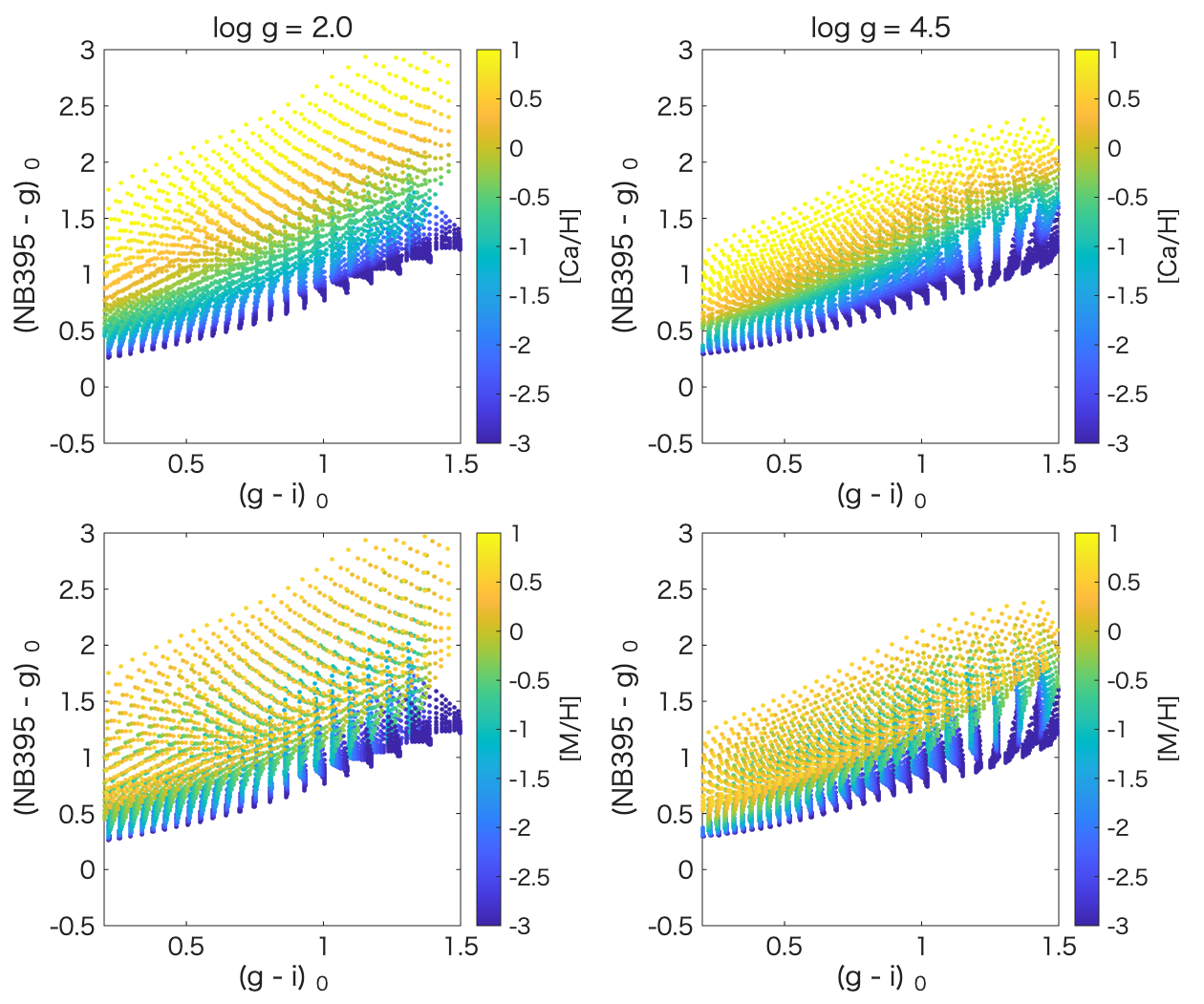}
\caption{Color-color diagrams illustrating the sensitivity of the NB395-based colors to [Ca/H] and [M/H]\@.
The left column presents MARCS model predictions with $\log g = 2$, representative of giant stars, while the right column shows models with $\log g = 4.5$, representative of dwarf stars.}
\label{Model_1}
\end{figure}

Color -- color diagrams based on [Ca/H] and [M/H] are shown in Figure~\ref{Model_1}. To minimize the effects of surface gravity, we adopt MARCS model points with $\log g = 2$ and $\log g = 4.5$, representative of giants and dwarfs, respectively. We find that, within the range $-1 \lesssim [{\rm M/H}] \lesssim -0.5$, model points with the same [M/H] largely overlap in the color -- color plane, whereas this degeneracy is not present when [Ca/H] is used. This behavior indicates that variations in [Ca/Fe] within a mixed disk and halo stellar population can introduce significant biases in metallicities inferred as [M/H] (the metallicity parameter provided by the MARCS models, approximately equivalent to [Fe/H]) from such color-color diagrams.

\begin{figure}[htbp!]
\includegraphics[width=90mm]{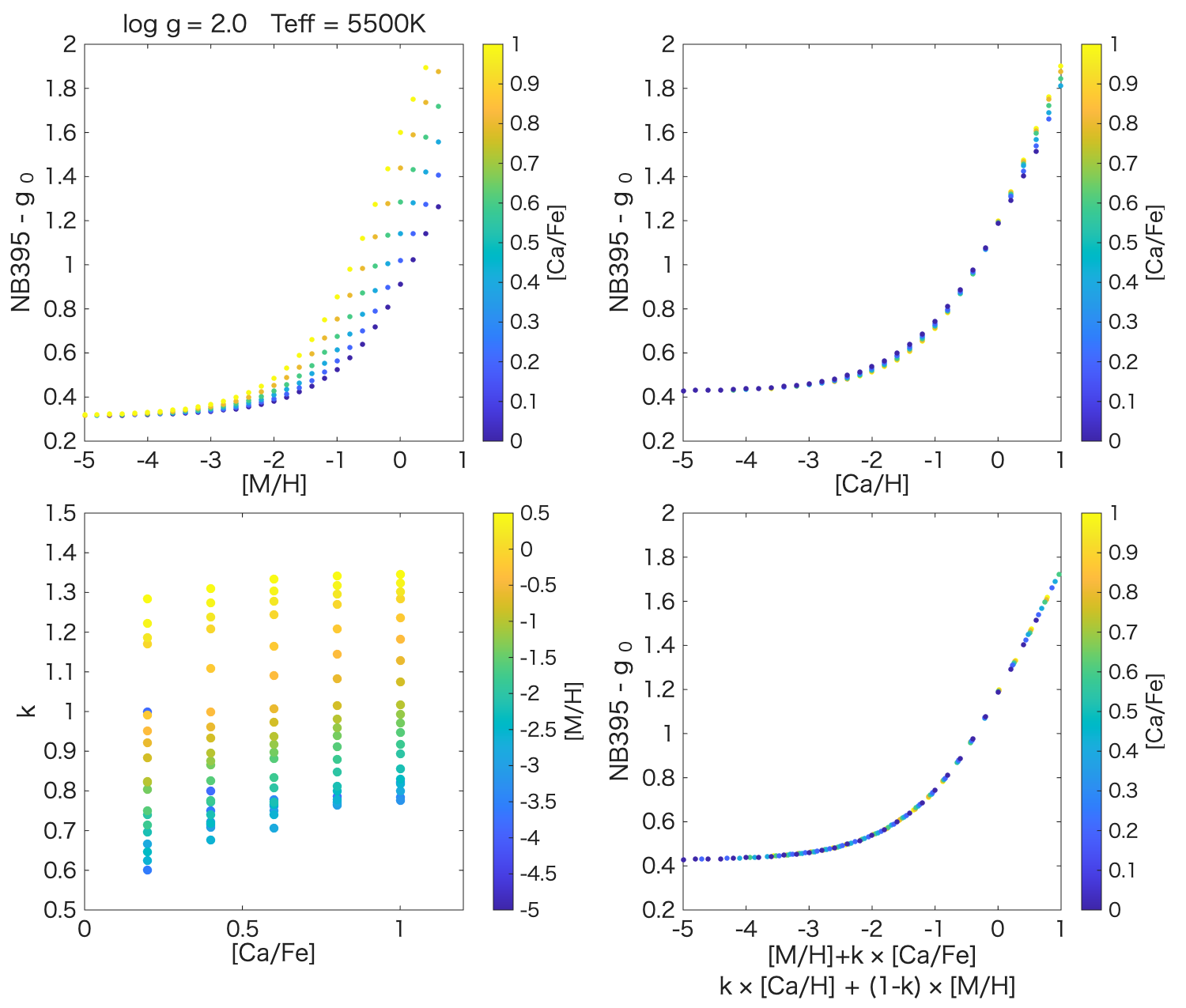}
\caption{Effect of [Ca/Fe] and [M/H] on the $(NB395 - g)_0$ color in a representative case.}
\label{Model_2}
\end{figure}

Nevertheless, [Ca/H] is not a perfect abundance indicator. We examine a representative case of giant stars with $\log g = 2$, $T_{\rm eff} = 5500$K ($(g-i)_0 \sim 0.5-0.6$), and $[{\rm Ca/Fe}] \ge 0$, as shown in Figure~\ref{Model_2}. At fixed reference color $(g-i)_0$, the abundance-sensitive color $(NB395 - g)_0$ does not vary monotonically with [M/H] and is strongly affected by variations in [Ca/Fe]. In contrast, $(NB395 - g)_0$ exhibits an approximately monotonic dependence on [Ca/H], although a weak sensitivity to [Ca/Fe] remains.
We therefore consider a composite chemical abundance indicator of the form $[{\rm M/H}] + k \times [{\rm Ca/Fe}]$ (an equivalent expression is $k \times [{\rm Ca/H}] + (1-k) \times [{\rm M/H}]$). 

We find that the characteristic value of $k$ varies under different conditions. For the present case, $k$ is approximately in the range $0.6$--$0.9$ in the metal-poor regime. 
When $k = 1$, this composite indicator reduces to $\mathrm{[Ca/H]}$. 
In the metal-rich regime, however, $k$ becomes larger than unity. 
This arises because, under the abundance-pattern assumptions adopted in the current model, 
the variations in $\alpha$-element abundances contribute to the response of the indicator and are therefore absorbed into the inferred value of $k$, leading to an apparent overestimation. 
The variation of $k$ thus reflects the fact that, when used as a metallicity indicator, $(NB395 - g)_0$ remains sensitive to multiple elemental abundances, although Ca still provides the dominant contribution.

\setcounter{secnumdepth}{0}
\setcounter{table}{0}   
\setcounter{figure}{0}
\renewcommand{\thetable}{B\arabic{table}}
\renewcommand{\thefigure}{B\arabic{figure}}

\section{B. Empirical Relation for the RS82 Standard Stars}
\begin{figure}[htbp!]
\includegraphics[width=90mm]{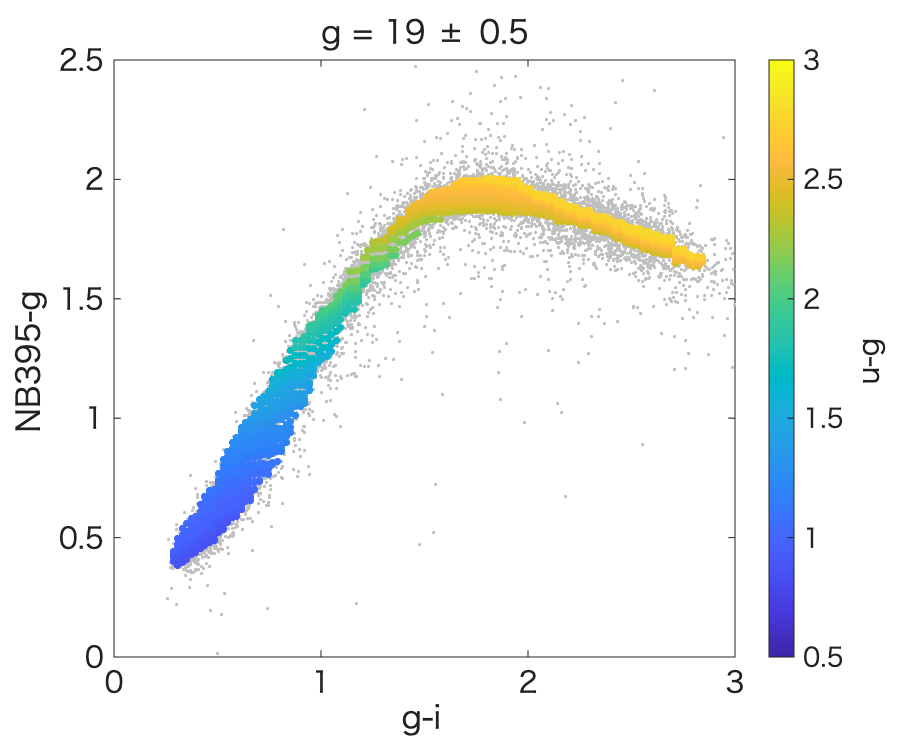}
\caption{The final empirical relation $NB395 = f(u-g, g-i, g)$ derived using the LSBoost regression, shown at a representative magnitude of $g=19$. The gray points show the training sample, while the colored points show the LSBoost predictions, color-coded by $u-g$.}
\label{RS82_relation}
\end{figure}

Figure~\ref{RS82_relation} illustrates the final empirical relation $NB395 = f(u-g, g-i, g)$ derived using the LSBoost regression after the iterative fitting procedure described in Section~4.3. The relation is shown at a representative magnitude of $g=19$. 

\setcounter{secnumdepth}{0}
\setcounter{table}{0}   
\setcounter{figure}{0}
\renewcommand{\thetable}{C\arabic{table}}
\renewcommand{\thefigure}{C\arabic{figure}}

\section{C. Variation of the Spatial Structure}
\begin{figure*}[htbp!]
\includegraphics[width=180mm]{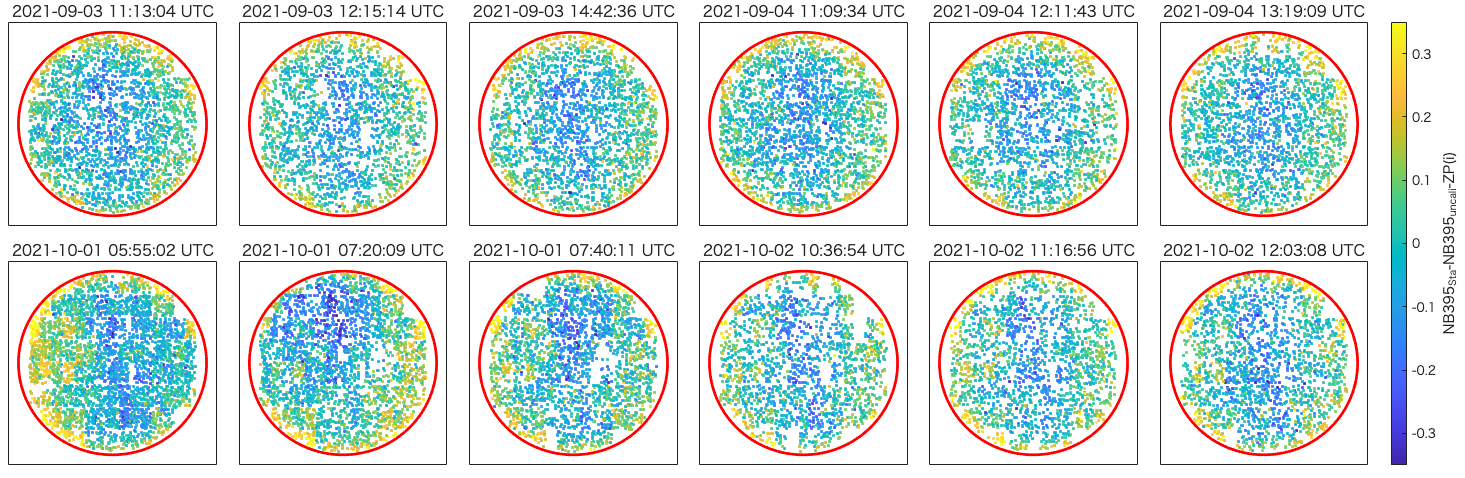}
\caption{Examples of the spatial structures during the four observing nights. The $X$--$Y$ spatial layout of each panel is the same as that in the upper-left panel of Figure~\ref{SS}.}
\label{SSvar}
\end{figure*}

Figure~\ref{SSvar} shows three examples of the spatial structures from each of the four observing nights. The spatial structures evolve only gradually throughout the nights of September 3, September 4, and October 2. In contrast, more rapid variations are seen on October 1, when the observing conditions were poorer. In addition to these variations within individual nights, the spatial structures observed in October differ noticeably from those in September. These results indicate that the spatial structure can vary both between observing runs and on shorter timescales, motivating our determination of the spatial correction independently for each visit.

\bibliographystyle{aasjournalv7.1}

\vspace{7mm} \noindent {\bf Acknowledgments}
B.H. thanks Prof. Haining Li (\CJKCN{李海宁}) for facilitating the contact and making possible the visit and collaborations.
B.H. acknowledges support from the National Natural Science Foundation of China (NSFC) under grant No. 124B2055, the National Key R\&D Program of China under grant No. 2024YFA1611601 and the Postdoctoral Fellowship Program of the China Postdoctoral Science Foundation (CPSF) under grant No. GZB20260765.
N.T. acknowledges support from JSPS KAKENHI grant Nos. 23H04894, 24K00682, 25K24630, 25K01046, 25H02196, 25H00674, and 26K21726.
M.C. acknowledges support from JSPS KAKENHI grant Nos. JP17H01101, JP24K00669, and JP25K24687. He especially acknowledges JP17H01101 for funding the HSC/NB395 filter.
H.Y. acknowledges support from the NSFC through projects No. 12222301 and 12173007, and from the National Key R\&D Program of China under grant 2024YFA1611901.
K.X. acknowledges support from the NSFC under grant No. 12403024; the CPSF under grant No. GZB20240731; the Young Data Scientist Project of the National Astronomical Data Center; and the China Postdoctoral Science Foundation under grant No. 2023M743447.
R.F.G.W. acknowledges the receipt of a JSPS Invitational Fellowship (No.~S26010), which facilitated her participation in this collaboration. She is grateful to  Schmidt Sciences for their support. 
E.N.K. and M.C. gratefully acknowledge support from Notre Dame Global Asia Research Collaboration Grants.
T.I. has been supported by IAAR Research Support Program in Chiba University Japan.
Y.S. acknowledges support from the Data-Scientist-Type Researcher Training Project of the Graduate University for Advanced Studies, SOKENDAI.

The Hyper Suprime-Cam (HSC) collaboration includes the astronomical communities of Japan and Taiwan, and Princeton University. The HSC instrumentation and software were developed by the National Astronomical Observatory of Japan (NAOJ), the Kavli Institute for the Physics and Mathematics of the Universe (Kavli IPMU), the University of Tokyo, the High Energy Accelerator Research Organization (KEK), the Academia Sinica Institute for Astronomy and Astrophysics in Taiwan (ASIAA), and Princeton University. Funding was contributed by the FIRST program from the Japanese Cabinet Office, the Ministry of Education, Culture, Sports, Science and Technology (MEXT), the Japan Society for the Promotion of Science (JSPS), Japan Science and Technology Agency (JST), the Toray Science Foundation, NAOJ, Kavli IPMU, KEK, ASIAA, and Princeton University. 

This paper makes use of software developed for the Large Synoptic Survey Telescope. We thank the LSST Project for making their code available as free software at \url{http://dm.lsst.org}.

This paper is based [in part] on data collected at the Subaru Telescope and retrieved from the HSC data archive system, which is operated by the Subaru Telescope and Astronomy Data Center (ADC) at National Astronomical Observatory of Japan. Data analysis was in part carried out with the cooperation of Center for Computational Astrophysics (CfCA), National Astronomical Observatory of Japan. The Subaru Telescope is honored and grateful for the opportunity of observing the Universe from Maunakea, which has the cultural, historical and natural significance in Hawaii.

This work has made use of data from the European Space Agency (ESA) mission {\it Gaia} (\url{https://www.cosmos.esa.int/gaia}), processed by the Gaia Data Processing and Analysis Consortium (DPAC, \url{https:// www.cosmos.esa.int/web/gaia/dpac/consortium}). Funding for the DPAC has been provided by national institutions, in particular the institutions participating in the Gaia Multilateral Agreement. 

This work made use of the data from LAMOST (Large Sky Area Multi-Object Fiber Spectroscopic Telescope, also known as the Guoshoujing Telescope) (\url{https://cstr.cn/31118.02.LAMOST}). LAMOST is a Chinese national mega-science facility, operated by National Astronomical Observatories, Chinese Academy of Sciences.

The Pan-STARRS1 Surveys (PS1) and the PS1 public science archive have been made possible through contributions by the Institute for Astronomy, the University of Hawaii, the Pan-STARRS Project Office, the Max Planck Society and its participating institutes, the Max Planck Institute for Astronomy, Heidelberg, and the Max Planck Institute for Extraterrestrial Physics, Garching, The Johns Hopkins University, Durham University, the University of Edinburgh, the Queen’s University Belfast, the Harvard-Smithsonian Center for Astrophysics, the Las Cumbres Observatory Global Telescope Network Incorporated, the National Central University of Taiwan, the Space Telescope Science Institute, the National Aeronautics and Space Administration under grant No. NNX08AR22G issued through the Planetary Science Division of the NASA Science Mission Directorate, the National Science Foundation grant No. AST-1238877, the University of Maryland, Eotvos Lorand University (ELTE), the Los Alamos National Laboratory, and the Gordon and Betty Moore Foundation.

{}

\end{CJK*}
\end{document}